\documentclass[twocolumn,showpacs,preprintnumbers,amsmath,amssymb,floatfix,nofootinbib]{revtex4}

\usepackage[breaklinks=true,colorlinks=true,backref,pagebackref]{hyperref} 

\usepackage{graphicx}
\usepackage{dcolumn}
\usepackage{bm}

\usepackage{graphicx,color}

\def\a{\alpha}
\def\b{\beta}
\def\g{\gamma}
\def\d{\delta}

\def\vt{\vartheta}\def\vta{\vartheta}

\def\lapl{ {\,  {\circ}\!-\!\!\! -{\bullet}\,}  }

\begin{document}

\title{Poincar\'e gauge theory of gravity: Friedman cosmology\\
  with even and odd parity modes. Analytic part}

\author{Peter Baekler} \email{peter.baekler@fh-duesseldorf.de}
\affiliation{Department of Media, Fachhochschule D\"usseldorf,
  University of Applied Sciences, 40474 D\"usseldorf, Germany}

\author{Friedrich W.\ Hehl}\email{hehl@thp.uni-koeln.de (corresp. author)}
\affiliation{Institute for Theoretical Physics, University of
  Cologne, 50923 K\"oln, Germany, \\
  \centerline{and}\\
  Department of Physics and Astronomy, University of Missouri,
  Columbia, MO 65211, USA}

\author{James M.\ Nester} \email{nester@phy.ncu.edu.tw}
\affiliation{Department of Physics \& Institute of Astronomy \& Center
  of Mathematics and Theoretical Physics\\ National Central
  University, Chungli 320, Taiwan, ROC\\}

\date{{06 November 2010}, {\it file
    PGCosmology/{PGcosmologyOne40X.tex}} }

\begin{abstract}
  We propose a cosmological model in the framework of the Poincar\'e
  gauge theory of gravity (PG). The gravitational Lagrangian is
  quadratic in curvature and torsion. In our specific model, the
  Lagrangian contains (i) the curvature scalar $R$ and the curvature
  pseudo-scalar $X$ linearly and quadratically (including an $RX$ term)
  and (ii) pieces quadratic in the torsion {\it vector} $\cal V$ and the
  torsion {\it axial} vector $\cal A$ (including a ${\cal V}{\cal A}$
  term). We show generally that in quadratic PG models we have nearly
  the same number of parity conserving terms (`world') and of parity
  violating terms (`shadow world'). This offers new perspectives in
  cosmology for the coupling of gravity to matter and antimatter. Our
  specific model generalizes the fairly realistic `torsion
  cosmologies' of Shie-Nester-Yo (2008) and Chen et al.\ (2009). With
  a Friedman type ansatz for an orthonormal coframe and a Lorentz
  connection, we derive the two field equations of PG in an explicit
  form and discuss their general structure in detail. In particular,
  the second field equation can be reduced to first order ordinary
  differential equations for the curvature pieces $R(t)$ and
  $X(t)$. Including these along with certain relations obtained from
  the first field equation and curvature definitions, we present a
  first order system of equations suitable for numerical
  evaluation. This is deferred to the second, numerical part of this
  paper.
\end{abstract}


\maketitle

\section{Introduction}\label{Sec.I}

In order to accommodate the {\it local Poincar\'e group} in spacetime,
Sciama \cite{Sciama62} and Kibble \cite{Kibble61} had to extend the
Riemannian spacetime of general relativity (GR) to a Riemann-Cartan
spacetime with non-vanishing torsion $T^\a$ (for the notation see the
end of the Introduction). Thereby the orthonormal coframe $\vt^\a$ and
the Lorentz connection $\Gamma^{\a\b}=-\Gamma^{\b\a}$ became
independent gauge potentials of weak and strong gravity,
respectively. The corresponding gauge field strengths are torsion
$T^\a=D\vt^\a$ and curvature $R^{\a\b}\sim D\Gamma^{\a\b}$, as spelled
out in Sec.~\ref{Sec.II}. There we also display the irreducible
decompositions of $T^\a$ and $R^{\a\b}$.

If one allows in a Yang-Mills manner for a gravitational Lagrangian
$V$ that is {\it quadratic in torsion and curvature,} we speak of a
Poincar\'e gauge theory of gravity (PG)
\cite{Erice79,Hayashi:1979wj,NesterCan,Egg,Erice95,Blagojevic,Ortin,Lewis}. In
Sec.~\ref{Sec.IIIA} we introduce the gravitational excitations
$H_\a=-\partial V/\partial T^\a$ and $H_{\a\b}=-\partial V/\partial
R^{\a\b}$ and recapitulate the general form (\ref{first}) and
(\ref{second}) of the two field equations of gravity.

Then, in Sec.~\ref{Sec.IIIB}, we turn to the conventional parity
conserving quadratic Lagrangian $V_+$, which includes the somewhat
degenerate Einstein-Cartan Lagrangian $V_{\text{EC}}$. Because of the
existence of the Euler 4-form of the curvature, we can show that one
curvature square piece is trivial. In Secs.~\ref{Sec.IIIC} and
~\ref{Sec.IIID}, we review parity violating {\it admixtures} to the
EC-Lagrangian that have been formulated in the past by different
groups, stressing in Sec.~\ref{Sec.IIIE} the importance of the
corresponding cosmological models of Shie-Nester-Yo \cite{Shie:2008ms}
and Chen et al.\ \cite{exSNY}.

Having in this way the PG at our disposal, we open for it in
Sec.~\ref{Sec.IV} a new `window' to a `shadow world': In
Sec.~\ref{Sec.IVA}, we show in a systematic way that, besides the
parity conserving Lagrangian $V_+$, there exists an equally important
Lagrangian $V_-$ the pieces of which are all {\it parity violating.}
Accordingly, for PG we propose the gravitational Lagrangian
$V_\pm=V_++V_-$. An equivalent Lagrangian has already been discussed
earlier by Obukhov et al.\ \cite{YuriEtAl}.

Because of the complexity of this general Lagrangian, we select for
further study in Sec.~\ref{Sec.IVB} in Eq.(\ref{L_BHN'}) the simpler
{\it 9-parameter Lagrangian} $V_{\rm BHN}$, which should carry the
characteristic features of parity conserving and parity violating
effects. In Sec.~\ref{Sec.IVC} a novel method is proposed for
diagonalizing the quadratic pieces in $V_{\rm BHN}$. No linearization
is involved and the output consists of exact analytic results. Besides
the Einstein mode $2^+$, we find for the {\it torsion modes} spin and
parity $0^\pm,1^\pm$ and for the {\it curvature modes} $0^\pm$. 

We calculate the gravitational excitations of $V_{\rm BHN}$
(Sec.~\ref{Sec.IVD}) and display in Sec.~\ref{Sec.IVE} the
corresponding field equations explicitly. They turn out to be first
order partial differential equations in torsion and curvature,
respectively.

We continue by looking closer into the structure of $V_{\rm BHN}$ and
its field equations. In Sec.~\ref{Sec.IVF} the Nieh-Yan identity is
used to show that the coupling constants of $V_{\rm BHN}$ occur only
in certain linear combinations in the field equations.

Eventually, in Sec.~\ref{Sec.V}, we turn to a cosmological model. In
Sec.~\ref{Sec.VA} the coframe and the torsion are assumed to be
homogeneous and isotropic in accordance with a
Friedman-Lema\^{\i}tre-Robertson-Walker (FLRW) type model and in
Sec.~\ref{Sec.VB} the corresponding irreducible pieces of the
curvature are calculated. We define a spinless perfect fluid in
Sec.~\ref{Sec.VC} and find then, in Sec.~\ref{Sec.VD}, the field
equations of gravity for this cosmological model. The first field
equation yields equations for the density $\rho(t)$ and for the
pressure $p(t)$ of the perfect fluid. These equations are subsequently
manipulated in order to bring them into a more transparent form. The
second field equation has also two independent components, namely
first order ordinary differential equations for $R(t)$ and $X(t)$. We
uncouple them and bring them in the very compact form
(\ref{sec_nonlinear}) and (\ref{sec_nonlinear1}) by introducing
certain `frequencies' $\omega_0,\omega_1,\omega_2,\omega_3$.

Now we are able to evaluate our exact results by numerical
methods. This will be done in follow up work.

\subsubsection*{Notation}
Our notation is as follows (see \cite{PRs,Birkbook}): We use the
formalism of exterior differential forms. We denote the {\it frame} by
$e_\a$, with the anholonomic or frame indices
$\a,\b,\dots=0,1,2,3$. Decomposed with respect to a {\it natural
  frame} $\partial_i$, we have $e_\a=e^i{}_\a\,\partial_i$, where
$i,j,\dots=0,1,2,3$ are holonomic or coordinate indices. The frame
$e_\a$ is the vector basis of the tangent space at each point of the
4D spacetime manifold. The symbol $\lrcorner$ denotes the interior and
$\wedge$ the exterior product. The {\it coframe} $\vt^\b=e_j{}^\b
dx^j$ is dual to the frame, i.e., $e_\a\lrcorner\vt^\b=\delta^\b_\a$.
The $^\star$ denotes the Hodge star operator that acts on the
quantities on its right, as, for instance, in $^\star(\Sigma_a\wedge
\vt^\b)$. If $\vt^{\a\b}:=\vt^\a\wedge\vt^\b$, etc., then we can
introduce the {\it eta-basis} by $\eta:=\,^\star 1$,
$\>\eta^\a:=e^\a\lrcorner\eta=\,^\star\vt^\a$,
$\>\eta^{\a\b}:=e^\b\lrcorner\eta^{\a}=\,^\star\vt^{\a\b}$, etc..
Parentheses surrounding indices $(\a\b) :=(\a\b+\b\a)/2$ denote
symmetrization and brackets $[\a\b] :=(\a\b-\b\a)/2$
antisymmetrization.

The coframe $\vt^\a$ and the $\eta$-system are related by
\begin{eqnarray}
\vta^\a\wedge\eta_\b &=& \delta^\a_\b\,\,\eta\,,\nonumber\\
  \vta^\a\wedge\eta_{\b\g}&=&
  \delta^\a_\g\,\,\eta_\b-\delta^\a_\b\,\,\eta_\g\,,\nonumber\\
  \vta^\a\wedge\eta_{\b\g\delta}&=&
  \delta^\a_\delta\,\,\eta_{\b\g}+\delta^\a_\g\,\,\eta_{\delta\b}
  +\delta^\a_\b\,\,\eta_{\g\delta}\,,\\
 \vta^\a\,\eta_{\b\g\delta\mu}&=&\delta^\a_\mu\,\,\eta_{\b\g\delta}
  -\delta^\a_\delta\,\,\eta_{\b\g\mu}+\delta^\a_\g\,\,\eta_{\b\delta\mu}
  -\delta^\a_\b\,\,\eta_{\g\delta\mu}\,.\nonumber
 \end{eqnarray}

 Differentiating the $\eta$'s, we find in a metric-affine space (for a
 definition see Sec.~\ref{Sec.IIB}) the relations
\begin{eqnarray}
  D\eta_{\alpha}& = & -2Q\wedge \eta_{\alpha}
  + T^\mu \wedge \eta_{\alpha\mu}\,,\nonumber\\
  D\eta_{\alpha\beta}& = & -2Q\wedge \eta_{\alpha\beta} +T^\mu \wedge
  \eta_{\alpha\beta\mu}\;,\nonumber\\
  D\eta_{\alpha\beta\gamma}&= & -2Q\wedge\eta_{\alpha\beta\gamma}+
  T^\mu\wedge\eta_{\alpha\beta\gamma\mu}\;, \nonumber\\
  D\eta_{\alpha\beta\gamma\delta}&
  = & -2Q\wedge\eta
  _{\alpha\beta\gamma\delta}\;,\label{Deta}
\end{eqnarray}
where $Q$ is the Weyl covector and $T^\mu$ the torsion.

We use the abbreviations: GR = general relativity theory (Riemann
spacetime), EC = Einstein-Cartan theory of gravity (Riemann-Cartan
spacetime with torsion and curvature, gauge Lagrangian linear in
curvature), PG = Poincar\'e gauge theory of gravity (Riemann-Cartan
spacetime, gauge Lagrangian arbitrary function of torsion and
curvature; often a quadratic function), MAG = metric-affine theory of
gravity (metric-affine spacetime, gauge Lagrangian arbitrary function
of in torsion, nonmetricity, and curvature; often a quadratic
function).

\section{\label{Sec.II}Spacetime of gauge theory of gravity}

\subsection{The coframe $\vt^\a$ and weak gravity}\label{Sec.IIA}

One of the fundamental structures in a gauge theory of gravity is the
{coframe} field $\vt^\a$---and all quantities are referred to
it. It is represented by four linearly independent 1-forms $\vt^\a$,
with $\a=0,1,2,3$. They can be decomposed with respect to a natural
coframe $dx^i$ according to $\vt^\a=e_i{}^\a dx^i$. Here the
$e_i{}^\a$ are the coordinate components of $\vt^\a$, also called
tetrad components.

Since a Riemannian {\it metric} $g$ with Lorentz signature is
assumed to exist, the coframe can always chosen to be orthonormal
according to
\begin{equation}\label{ortho}
g=g_{ij}\,dx^i\otimes dx^j=g_{\a\b}\,\vt^\a\otimes\vt^\b\,,
\end{equation}
with $g_{\a\b}=\text{diag}(-1,1,1,1)$. In this way the metric is
absorbed by the coframe and has no longer independent physical degrees
of freedom.

This choice is convenient and will be kept throughout this
article. However, sometimes one may want to choose arbitrary coframes:
then the metric emerges explicitly again. Also for that reason, the
metric and the coframe, besides the linear connection (see
Sec.~\ref{Sec.IIB}), are treated in the variational principle as
independent gauge field variables. However, it eventually turns out
that the field equations resulting {}from the variation of the metric
and of the coframe are equivalent. In other words, the orthonormal
``gauge'' of the coframe, which we use in this article, doesn't
restrict the generality of our considerations.

We call $\vt^\a$ the potential of {\it weak} gravity of the
{\it Newton-Einstein} type. It couples to matter via the Einstein
gravitational constant $\kappa$, which has the dimension of a
reciprocal force. Basically, $\vt^\a$ represents four gauge
boson fields---each of helicity 1---that conspire, at least in linear
approximation, to build up the massless spin 2 graviton modes of
general relativity (GR), provided the appropriate Hilbert Lagrangian
is chosen.

\subsection{The linear connection $\Gamma_\a{}^\b$ and the hypothesis
  of strong gravity}\label{Sec.IIB}

In classical gauge theories of gravity the linear connection of
spacetime $\Gamma_\a{}^\b$, besides the coframe $\vt^\a$, is assumed
to exist as an independent field variable. This is a physical
hypothesis that has eventually to be checked by experiment. We call
$\Gamma_\a{}^\b$ the potential of {\it strong} gravity of {\it
  Yang-Mills} type. It is represented by $4\times 4=16$ bosonic
one-forms fields that can be decomposed according to
$\Gamma_\a{}^\b=\Gamma_{i\a}{}^\b dx^i$. They couple to matter in a
Yang-Mills like fashion via a hypothetical coupling
constant $\varrho$ of the dimension (action)$^{-1}$.

A differential manifold equipped with a metric $g$ and a linear
connection $\Gamma_\a{}^\b$ is called a metric-affine space. If the
linear connection is unconstrained, then $\Gamma_\a{}^\b$ has values
in the Lie algebra of the general linear group $GL(4,R)$. The gauge
theory with independent metric and independent connection is called
{\it metric-affine} (gauge theory of) {\it gravity} (MAG). Since in
MAG the connection components $\Gamma_{i\a}{}^\b$ carry three indices,
the strong gravity potential can provide additional strong gravity
modes of up to spin 3 (see \cite{Baekler:2006vw}).

\subsection{Beyond general relativity: relaxing the torsion}\label{Sec.IIC}

Let us start {}from GR and assume the connection to be Riemannian,
namely $\widetilde{\Gamma}_\a{}^\b$; we will always indicate the
Riemannian nature of a quantity by a tilde. Then
$\widetilde{\Gamma}_\a{}^\b$ doesn't provide a mode that is
independent of the coframe. With a suitable Lagrangian, this
corresponds to GR.

If the connection is metric-compatible but non-Riemannian, it carries
an independent piece that has values in the Lie-algebra of the Lorentz
group $SO(1,3)$. In orthonormal frames, we have for the {\it Lorentz-}
(or spin-)\\{\it connec\-tion} the relation
$\Gamma^{\a\b}=-\Gamma^{\b\a}$. It represents 6 bosonic one-form
fields for strong gravity. Its maximum spin is 2. The independent fields,
coframe $\vt^\a$ and Lorentz-connection $\Gamma^{\a\b}$, represent the
gauge potentials of the Poincar\'e group. The corresponding gauge
field theory is called the Poincar\'e gauge theory of gravity (Poincar\'e
gravity or PG).

In PG, the {\it contortion}
\begin{equation}\label{contortion0}
K^{\a\b}:=\widetilde{\Gamma}^{\a\b}-{\Gamma}^{\a\b}=-K^{\b\a}
\end{equation}
measures the difference between the Riemann and the Riemann-Cartan
(RC) geometry and, as a difference between two connections, it
constitutes a tensor. Alternatively, the deviation {}from the Riemannian
geometry of GR can be described by the {\it torsion}
\begin{equation}\label{torsion}
  T^\a:=\stackrel{\Gamma}{D}\vt^\a=d\vt^\a+\Gamma_\b{}^\a\wedge\vt^\b
  =\frac 12 T_{ij}{}^\a dx^i\wedge dx^j\,.
\end{equation}
Here $\stackrel{\Gamma}{D}$ is the exterior covariant derivative with
respect to the connection $\Gamma_\a{}^\b$. The newly emerging
Lorentz-connection modes reflect themselves also in those of the
torsion, since the second term in the torsion,
$\Gamma_\b{}^\a\wedge\vt^\b$, depends on $\Gamma_\a{}^\b$. It can be
shown (see \cite{PRs}) that torsion and contortion are related by
\begin{eqnarray}\label{torcontor}
  T^\a=\vt^\b\wedge K_\b{}^\a,\;\;  K_{\alpha\beta}=
  e_{[\alpha}\lrcorner T_{\beta ]} -{1\over 2}\,
  ( e_{\alpha}\lrcorner e_{\beta}\lrcorner T_{\gamma})\vartheta^{\gamma}.
\end{eqnarray}

On the level of the gauge {\it potentials,} we have then in PG the
frame $\vt^\a$ and the Lorentz-connection
$\Gamma^{\a\b}=-\Gamma^{\b\a}$. On the level of the newly introduced
torsion, the translation gauge {\it field strength,} we can execute
an irreducible decomposition in order to learn more about its
structure. It can be decomposed according to $24=16\oplus 4\oplus 4$
into three pieces: into a second rank tensor piece ({tentor}),
$^{(1)}T^\a$, into a vector piece, namely, in components, the trace of
the torsion ({trator}),
\begin{equation}\label{trator}
  ^{(2)}T^\a:=-\frac 13\, {\cal V}\wedge\vt^\a\quad\text{with}\quad
   {\cal V}:=e_\b\lrcorner T^\b\,,
\end{equation}
and into an axial vector piece ({axitor}), which corresponds in
components to the totally antisymmetric piece of the torsion (the star
denotes the Hodge operator):
\begin{equation}\label{axitor}
  ^{(3)}T^{\alpha} =  \frac{1}{3}\,
  ^{\star}\left({\cal A}\wedge\, {\vartheta}^{\alpha}\right)\quad
  \text{with}\quad {\cal A}:=\,
  ^{\star}\left( {\vartheta}_{\alpha}\wedge T^{\alpha}\right)\,.
\end{equation}
We have then the irreducible decomposition
\begin{equation}\label{irrtorsion}
  T^\a=\,\underbrace{^{(1)}T^\a}_{\text{tentor}}\,+\,
  \underbrace{^{(2)}T^\a}_{\text{trator}}
  \,+\,\underbrace{^{(3)}T^\a}_{\text{axitor}}\,.
\end{equation}
The tensor piece can carry at most spin 2 modes, whereas the vector
and the axial vector pieces are good for at most spin 1 modes.
\medskip

\subsection{Keeping the nonmetricity to zero}\label{Sec.IID}

In MAG, besides the torsion $T^\a$, we have a nonvanishing
nonmetricity
\begin{equation}\label{nom}
  Q_{\a\b}:=-\stackrel{\Gamma}{D}g_{\a\b}=Q_{\b\a}
  =-dg_{\a\b}+2\Gamma_{(\a\b)}\,.
\end{equation}
Therefore, the nonmetricity $Q_{\a\b}$ of the metric-affine geometry
is a measure for the difference between the linear connection
$\Gamma_\a{}^\b$ and the Lorentz-connection. In an orthonormal coframe
$\vt^\a$ the metric referred to the coframe $g_{\a\b}$ is a constant
and the modes of the symmetric\footnote{Not to be confused with the
  symmetric part of the connection
  $\Gamma_{(\a\b)\g}=e_{(\a}\lrcorner\Gamma_{\b)\g}$.}
$\Gamma_{(\a\b)}=\Gamma_{i(\a\b)}dx^i$ are passed through, according
to (\ref{nom}), to the nonmetricity $Q_{\a\b}=Q_{i\a\b}dx^i$, with the
components $Q_{i\a\b}=Q_{i\b\a}$. There emerge, besides the 6 of the
Lorentz-connection, 10 more bosonic one-form fields $Q_{\a\b}$ clearly
encompassing strong gravity contributions of spin 0,1,2, and 3.

In MAG, we have the gravitational potentials $\vt^\a$ and
$\Gamma_\a{}^{\b}$. The gauge field strength attached to
the coframe $\vt^\a$ is the torsion (\ref{torsion}), that attached to
the linear connection $\Gamma_\a{}^{\b}$ the curvature 2-form
\begin{equation}\label{corv}
  R_\a{}^{\b}:=d\Gamma_\a{}^\b-\Gamma_{\a}{}^{\g}\wedge\Gamma_{\g}{}^{\b}\,.
\end{equation}
We can raise the index $\a$ and can decompose the curvature into the
antisymmetric `rotational' piece and the symmetric `strain' piece
according to
\begin{eqnarray}\label{deccurv}
 && R^{\a\b}=W^{\a\b} + Z^{\a\b}\\ \nonumber
 && \text{with}\quad W^{\a\b}:=R^{[\a\b]}\,, \;Z^{\a\b}:=R^{(\a\b)}\,.
\end{eqnarray}

Even though we succeeded to relate a quadratic MAG-Lagrangian to a
consistent classical field theory of massless spin 3 fields via the
tracefree part of the nonmetricity \cite{Baekler:2006vw}, we will
restrict ourselves in this article to vanishing nonmetricity,
$Q_{\a\b}=0$; consequently the strain curvature vanishes, too:
$Z^{\a\b}=0$. {}From a phenomenological point of view the overwhelming
importance of the (rigid) Poincar\'e group in special relativity
directs our attention primarily to the gauge theory of the {\it local}
Poincar\'e group, namely PG. Accordingly, from now on $Z^{\a\b}=0$ and
$R^{\a\b}=W^{\a\b}$.

\subsection{Irreducible decomposition of the rotational curvature}\label{Sec.IIE}

For the physical interpretation it is significant to understand the
different pieces of the curvature. The rotational curvature decomposes
irreducibly into six pieces, see
\cite{WallnerU4,McCrea:1992wa,PRs,Obukhov:2006gea,Vassiliev:2003dk},
according to
\begin{eqnarray}\label{Wdecomp}
R_{\a\b}&=& \underbrace{^{(1)}R_{\a\b}}_{\text{weyl 10}}+
 \underbrace{^{(2)}R_{\a\b}}_{\text{paircom 9}}+
 \underbrace{^{(3)}R_{\a\b}}_{\text{pscalar 1}}\nonumber\\
&&\hspace{-9pt} +\underbrace{^{(4)}R_{\a\b}}_{\text{ricsymf 9}}+
 \underbrace{^{(5)}R_{\a\b}}_{\text{ricanti 6}}+
\underbrace{^{(6)}R_{\a\b}}_{\text{scalar 1}}\,.
\end{eqnarray}
The number of independent components is specified subsequent to the
(computer) name of the corresponding irreducible piece. Pseudoscalar
and scalar qualify as linear Lagrangians. We take {}from the literature
(remember that $\vt_{\a\b}=\vt_\a\wedge \vt_\b$ and
$\eta_{\a\b}=\,^\star\vt_{\a\b}$)
\begin{eqnarray}\label{irr3}
  ^{(3)}R_{\a\b}&=&-\frac{1}{12}\,X\eta_{\a\b}\,,\\ && \nonumber X:=e_\a\lrcorner
  X^\a\,,\quad X^\a:=\,^\star \left(R^{\b\a}\wedge \vt_\b \right)\,;\\
 \label{W6} ^{(6)}R_{\a\b}&=&-\frac{1}{12}\,R\vt_{\a\b}\,,\\
  \nonumber &&  R:=e_\a\lrcorner R^\a\,, \quad R^\a:=e_\b\lrcorner
  R^{\a\b}\,.
\end{eqnarray}
We recognize that the scalar $R$ and the pseudoscalar $X$ play a
preferred role. Note that $X$ is purely post-Riemannian, that is,
$\widetilde{X}\equiv 0$. In components we have
\begin{equation}\label{WXcomp}
  X=\eta_{\a\b\g\d}R^{[\a\b\g\d]}/4!\quad\text{and}\quad R=R_{\b\a}{}^{\a\b}\,,
\end{equation}
with the decomposition $R^{\a\b}=R_{\g\d}{}^{\a\b}\vt^{\g\d}/2$.

\section{Poincar\'e gauge theory of gravity (PG)}\label{Sec.III}

\subsection{Lagrangian and field equations}\label{Sec.IIIA}

In PG, we have the gravitational potentials $\vt^\a$ and
$\Gamma^{\a\b}=-\Gamma^{\b\a}$. The corresponding gauge field strengths are
the torsion $T^\a$ and the rotational (`Lorentz') curvature
\begin{equation}\label{curv}
  R^{\a\b}:=d\Gamma^{\a\b}-\Gamma^\a{}_\g\wedge\Gamma^{\g\b}=-R^{\b\a}\,.
\end{equation}
We assume a first order Lagrangian consisting of a gauge and a
minimally coupled matter part,
\begin{equation}\label{lagrangian}
  L=V\left(g_{\a\b},\vt^\a,T^\a, R^{\a\b}\right)
  +L_{\text{mat}}\left(g_{\a\b},\vt^\a,\Psi,\stackrel{\Gamma}{D}\Psi\right),
\end{equation}
with the matter field(s) $\Psi$. Then we can define the
translation and the Lorentz {\it excitations,} respectively,
\begin{equation}\label{excit}
  H_\a:=-\frac{\partial
    V}{\partial T^\a}\,,\quad  \, H_{\a\b}:=-\frac{\partial V}{\partial
    R^{\a\b}}=-H_{\b\a}\,,
\end{equation}
and the canonical matter currents of {\it energy-momentum} and {\it
  spin} (angular momentum) according to
\begin{equation}\label{mattercurrents}
\Sigma_\a:=\frac{\d L_{\text{mat}}}{\d\vt^\a}\,,\quad
\tau_{\a\b}:=\frac{\d L_{\text{mat}}}{\d\Gamma^{\a\b}}=-\tau_{\b\a}\,;
\end{equation}
in the case of a minimally coupled matter Lagrangian, as in
(\ref{lagrangian}), the variational derivatives degenerate to partial
derivatives.

The action principle yields the field equations \cite{Erice79}
\begin{eqnarray}\label{first}
  DH_\a -E_\a &=&\Sigma_\a\hspace{20pt}\text{(first)}\,,\\ \label{second}
DH_{\a\b}-E_{\a\b} &=&\tau_{\a\b}\hspace{15pt}\text{(second)}\,,
\end{eqnarray}
with the gauge currents of energy-momentum and spin
\begin{eqnarray}\label{gaugecurrents}
E_\a&\hspace{-3pt}:=\hspace{-3pt}& e_{\alpha}\lrcorner V + (e_{\alpha}\lrcorner
T^{\beta})\wedge H_{\beta} + (e_{\alpha}\lrcorner
R^{\b\gamma})\wedge H_{\beta\gamma}, \hspace{9pt}\\
E_{\a\b}&\hspace{-3pt}:=\hspace{-3pt}& -\vt_{[\a}\wedge H_{\b]}\,.\label{gc2}
\end{eqnarray}

If the gauge Lagrangian $V$ is prescribed {\it explicitly,} we can compute
first the excitations $H_\a,H_{\a\b}$ by partial differentiation of
$V$ and subsequently the gauge currents $E_\a,E_{\a\b}$ by
substitution; these quantities are then inserted into the two field
equations (\ref{first}),(\ref{second}). As noted already above, the
field equation resulting {}from a variation of the metric $g_{\a\b}$ is
equivalent to (\ref{first}), provided (\ref{second}) is fulfilled.

The matter currents on the right-hand-side of the field equations
(\ref{first}) and (\ref{second}) can be understood as those of a spin
fluid (see \cite{Halbwachs,hyperfluid,hyperfluidII}). An {\em
  approximate} representation of such a spin fluid can be specified as
follows: If the fluid moves with the velocity ${\mathbf u}={ u}^\a
e_\a$, that is, with the flow 3-form ${\cal U}:={\mathbf
  u}\lrcorner\eta={u}^\a\eta_\a$, and transports an energy-momentum
density $p_\a$ and a spin density $s_{\a\b}=-s_{\b\a}$, then a
convective Weyssenhoff ansatz for the matter currents reads
\begin{equation}\label{convective}
  \Sigma_\a=p_\a\, {\cal U}\quad \text{and}\quad
\tau_{\a\b}=s_{\a\b}\,{\cal  U}\,.
\end{equation}

\subsection{Quadratic Yang-Mills type Lagrangian with even parity
  terms}\label{Sec.IIIB}

A quadratic Lagrangian of PG of the Yang-Mills type has the
general structure ($\lambda_0$ is the cosmological constant)
\begin{eqnarray}\label{quadraticL}
  V \sim \frac{1}{\kappa}\left( \text{curv}
  +\text{torsion}^2+\lambda_0\right) + \frac{1}{\varrho}\,\text{curv}^2\,.
\end{eqnarray}
Since the coframe $\vt^\a$ and the Lorentz connection $\Gamma^{\a\b}$ are
independent variables, such a first order Lagrangian yields second
order field equations; {\it higher derivatives do not emerge.}

The {\it simplest} non-trivial Lagrangian corresponds to the first
term on the right-hand-side of (\ref{quadraticL}). It is of the
Hilbert type, i.e., linear in the curvature, namely the so-called
Einstein-Cartan Lagrangian, see (\ref{W6}),
\begin{equation}\label{VEC0}
  V_{\text{EC}}=\frac{1}{2\kappa}\eta_{\a\b}\wedge
  R^{\a\b}=\frac{1}{2\kappa}\eta_{\a\b}\wedge
  \,^{(6)}R^{\a\b}=\frac{1}{2\kappa}{}^\star R\,.
\end{equation}
The corresponding two field excitations, ${H_\a=0}$ and
${H_{\a\b}=-\eta_{\a\b}/(2\kappa)}$, if substituted into
(\ref{first}) and (\ref{second}), yield the field equations
\cite{Sciama62,Kibble61,Trautman}:
\begin{eqnarray}\label{1stEC}
  \frac 12 \eta_{\a\b\g}\wedge R^{\b\g} &=&\kappa \Sigma_\a\,, \\ \label{2ndEC}
  \frac 12 \eta_{\a\b\g}\wedge T^\g &=&\kappa\tau_{\a\b}\,.
\end{eqnarray}
The viable Einstein-Cartan(-Sciama-Kibble) theory (EC), as compared to
GR, supplies an additional spin-contact interaction of {\it weak}
gravitational origin, since only Einstein's gravitational constant
enters (\ref{1stEC}),(\ref{2ndEC}). The Lorentz connection
$\Gamma^{\a\b}$ cannot propagate and thus EC
represents\footnote{Recently, an EC-model with fermionic matter and
  its application to the early universe has been discussed by Ribas \&
  Kremer \cite{Ribas:2009yg} and Dolan \cite{Dolan:2009ni}.} a
degenerate PG. In order to enable $\Gamma^{\a\b}$ to propagate, we
have to use additionally at least the {\it quadratic} curvature piece
in (\ref{quadraticL}); for discussions on the physical relevance of
torsion one should also compare Shapiro \cite{Shapiro:2001rz} and Ni
\cite{Wei-Tou2010}.

%

Esser \cite{Esser}, see also \cite{Hehl:1999sb}, constructed the most
general quadratic Lagrangian with even ($+$) parity pieces (for this
notion see Sec.~\ref{Sec.IIIC}). For a RC-spacetime it reads:
\begin{eqnarray}\nonumber
  \label{QMA}V_{+}  &=&
  \frac{1}{2\kappa}(-a_0R^{\alpha\beta}\wedge\eta_{\alpha\beta}
    -2\lambda_{0}\eta\\  &&
    \hspace{20pt} +T^\alpha\wedge\textstyle\sum
    \limits_{I=1}^{3}a_{I}{}^{\star(I)}\nonumber
    T_\alpha)\\ & &\hspace{-8pt} -\frac{1}{2\varrho}R^{\alpha\beta}
  \wedge{}\textstyle\sum\limits_{I=1}^{6}w_{I}{}^{\star(I)}
  R_{\alpha\beta}\,;
\end{eqnarray}
without restricting the generality of our considerations, we can
choose $\varrho>0$.  In a RC-space, Esser found the additional term
\begin{equation}
-\frac{1}{2\varrho}\,R^{\alpha\beta} \wedge{}^\star \!
  \left[w_7\,\vartheta_\alpha\wedge(e_\gamma\lrcorner
    ^{(5)}R^\gamma{}_{\beta} )\right]\,.
\end{equation}
However, this term can be transformed successively into a pure $w_5$
term: Let us first consider invariants of the form
$R^{\alpha\beta}\wedge\, ^{\star}\left[ {\vartheta}_{\alpha}\wedge\,
  \left(e_{\gamma}\lrcorner\, ^{(A)}R^{\gamma}{}_{\beta}\right)\right]$
with $A\in \{1\cdots 6\}$. For $A=5$ we find
\begin{eqnarray}\label{identityw7}
&& R^{\alpha\beta}\wedge\, ^{\star}\left[
{\vartheta}_{\alpha}\wedge\, \left(e_{\gamma}\lrcorner\,
^{(5)}R^{\gamma}{}_{\beta}\right)\right]\nonumber\\ &&\hspace{20pt} = \,
^{(5)}R^{\alpha\beta}\wedge\, ^{\star}\left[
{\vartheta}_{\alpha}\wedge\, \left(e_{\gamma}\lrcorner\,
^{(5)}R^{\gamma}{}_{\beta}\right)\right]\nonumber\\ &&\hspace{20pt} = \,
^{(5)}R^{\a\b}\wedge\, ^{\star (5)}R_{\a\b}\,.
\end{eqnarray}
Thus we can absorb the $w_7$ term into the $w_5$ term. Consequently,
without restricting the generality of our considerations, we can put
$w_7=0$.

In our Lagrangian $V_{+}$ in (\ref{QMA}) not all the constants are
independent. In a Riemannian as well as in a RC-spacetime, the
integrand of the topological Euler 4-form
\begin{eqnarray}\label{Euler}
  B_{RR^{(*)}}&=&-{1\over 2}R_{\a}{}^{\b}\wedge R^{(*)}_{\b}{}^{\a}
  ={1\over 4}\eta_{\b\a\d\g }\,R^{\a\b}\wedge R^{\g\d}\nonumber\\ &=&
  dC_{RR^{(*)}}\, 
\end{eqnarray}
is exact, with
\begin{equation}
  C_{RR^{(*)}}={1\over 4}
  \eta^{\a\hspace{5pt}\g}_{\hspace{5pt}\b\hspace{5pt}\d}\,\Bigl(R_\a{}^\b\wedge
  \Gamma_\g{}^\d + {1\over 3}\Gamma_\a{}^\b\wedge\Gamma_\g{}^
  \varepsilon\wedge\Gamma_\varepsilon{}^\d\Bigr)\,.
\end{equation}
The dual is here taken with respect to the frame indices $\a,\b$ of
the curvature 2-form $R_\a{}^\b$, it has to be carefully distinguished
from the Hodge dual.  Then, together with the Bach-Lanczos identity
\cite{PRs}, Eq.(A.3.7),
\begin{eqnarray}
  R^{(\a\vert\g}\wedge R^{(*)}{}^{\vert\b)}{}_\g -\frac 14
  g^{\a\b}\,  R^{\mu\nu}\wedge R^{(*)}_{\mu\nu} =0\,,
\end{eqnarray} 
one can show that only five of the six $w_I$'s are linearly
independent.

The excitations can now be calculated by differentiation:
\begin{eqnarray}
  H_{\alpha}& =& -\frac{1}{\kappa}
    \textstyle\sum\limits_{I=1}^{3} a_{I}\, {^{\star(I)}T}_{\alpha}\,,
    \label{Ha-excit}\\ \label{Hab-excit}
    H_{\alpha\beta}&=&\frac{a_{0}}{2\kappa}{\eta}_{\alpha\beta}+
    \frac{1}{\varrho}\textstyle\sum\limits_{I=1}^{6}w_{I}
      \,^{\star (I)}R_{\alpha\beta}\, .
\end{eqnarray}
Because of (\ref{W6}), the last equation can be slightly rewritten as
\begin{equation}\label{H_ab0}
  H_{\a\beta}=\left(\frac{a_{0}}{2\kappa}- \frac{w_{6}}{12\varrho}\,R
  \right){\eta}_{\alpha\beta}+\frac{1}{\varrho} \textstyle\sum
  \limits_{I=1}^{5} w_{I}\, ^{\star (I)}R_{\alpha\beta}\, .
\end{equation}
There have been numerous investigations into the properties of the
Lagrangian (\ref{QMA}). In linear approximation, on a flat Minkowskian
background, Eq.\ (\ref{QMA}) encompasses, besides the weak gravity
modes of the coframe, propagating strong gravity modes of the
Lorentz-connection with spin $2^\pm$, $1^\pm$, and $0^\pm$, as shown
by Hayashi \& Shirafuji \cite{Hayashi:1979wj},
by Sezgin \& van Nieuwenhuizen \cite{Sezgin:1979zf}, and by Kuhfuss \&
Nitsch \cite{Kuhfuss:1986rb}. For a model with quadratic curvature
Lagrangian in which only the Lorentz-connection is dynamic, compare
Cho et al.\ \cite{Cho:2009fa}.

A good dynamic mode transports positive energy at speed $\le c$. At
most three modes can be simultaneously dynamic; all the cases were
tabulated; many combinations are satisfactory to linear order. The
Hamiltonian analysis, as shown by Blagojevi\'c and Nikoli\'c
\cite{Blagojevic:1983zz,Nikolic:1984xi}, revealed the related
constraints. In more detailed investigations
\cite{Hecht:1996ay,Chen:1998ad,Yo:1999ex,Yo:2001sy} it was concluded
that effects due to nonlinearities could be expected to render all of
these cases physically unacceptable, with the exception of two
``scalar'' connection modes with spin $0^+$ and spin $0^-$.

Before we come back to the mode analysis in Sec.~\ref{Sec.IVC}, we want to
extend the gravitational Lagrangian such that also odd parity pieces
are included.

\subsection{Even parity and odd parity Lagrangians, twisted and
  untwisted forms}\label{Sec.IIIC}

Let us study the spatial reflection or parity transformation; the sign
of the time axis will be kept fixed. A (pure) scalar field $\Phi(x)$
remains invariant under the parity transformation or if we
transform a right-handed coordinate system $x^i$ into a left-handed
one $x^{i'}$: $\Phi(x')=\Phi(x)$. In contrast, a twisted scalar field
$\hat{\Phi}(x)$ (also called a pseudoscalar field) changes its sign
under those circumstances, that is, the sign of the determinant
$J:=\det|| \partial x^i/\partial x^{i'}||$ of the Jacobian
transformation matrix enters its transformation law:
$\hat{\Phi}(x^{i'})= (\text{sign}\,J) \hat{\Phi} (x)$. The analogous
behavior characterizes the relation between twisted and untwisted
forms; for a mathematical discussion compare Frankel \cite{Ted}.

A {\em Lagrangian} 4-form $L$ has to be a {\it twisted} 4-form in
order to make its action $W:=\int_{\Omega_4}L$ a pure scalar. These
Lagrangians are also called even parity Lagrangians. However, in
physics we know since the discovery of parity violation in the weak
interaction in 1956, see Sozzi \cite{Sozzi} for a review, that also
odd parity would-be Lagrangians can occur; they have to be multiplied
by {\it pseudoscalar coupling constants} in order to transform them
to decent (twisted) Lagrangians, which can be added to the other
even parity Lagrangian pieces.

In PG, the field strengths $T^\a$ and $R^{\a\b}$ are untwisted
2-forms, similarly the potentials $\vt^\a$ and $\Gamma^{\a\b}$ are
untwisted 1-forms. One may compare the case of electrodynamics with
the untwisted potential $A$ and the untwisted field-strength $F=dA$
(the differential $d$ is untwisted). Consequently, a twisted
Lagrangian, according to (\ref{excit}), leads to the excitations
$H_\a$ and $H_{\a\b}$ being twisted 2-forms and the material currents
$\Sigma_\a$ and $\tau_{\a\b}$, see (\ref{mattercurrents}), being
twisted 3-forms.

The Einstein-Cartan Lagrangian
\begin{equation}\label{VEC}
  V_{\text{EC}}=\frac{1}{2\kappa}R^{\a\b}\wedge{}^{\star\!}\left(\vt_\a\wedge
    \vt_\b\right)=\frac{1}{2\kappa}{}^{\star} R
\end{equation}
is twisted, since the Hodge star in our formalism (see
\cite{Birkbook}, Sec.\ C.2.8) is twisted, that is, it maps twisted
into untwisted forms and vice versa. The Maxwell Lagrangian
\begin{equation}\label{VMax}
V_{\text{Max}}=-\frac{Y_0}{2}F\wedge\,^\star F
\end{equation}
is also twisted, hence of even parity, where $Y_0$ is the (scalar)
vacuum admittance. We recognize that an odd number of stars occurring
in a Lagrangian, which is expressed in terms of field strength and
potentials, guarantees its standard twisted nature. In contrast, the
topological Chern type Lagrangian
\begin{equation}\label{VMax'}
V_{\text{Max$'$}}=-\frac{Y_1}{2}F\wedge F
\end{equation}
is only twisted (even parity), if we declare $Y_1$ to be
a pseudoscalar; the analogous is true in gravity for
\begin{equation}\label{VEC'}
  V_{\text{EC$'$}}=\frac{1}{2\kappa'}R^{\a\b}\wedge
 \left(\vt_\a\wedge\vt_\b\right)=\frac{1}{2\kappa'}\,^{\star\!} X\,,
\end{equation}
with the pseudoscalar constant $\kappa'$; note that this Lagrangian
vanishes identically in a Riemannian space, since
$\widetilde{R}^{[\a\b\g\d]}=0$.

\subsection{Parity violating admixtures to the Einstein-Cartan
  Lagrangian}\label{Sec.IIID}

Already in 1964, Leitner and Okubo \cite{Leitner:1964tt} wondered
about possible odd parity terms in the gravitational
Lagrangian. Related questions were addressed by Hayashi
\cite{Hayashi:1973nf} and Hari Dass \cite{HariDass:1976kc}. The effect
of adding the non-Riemannian odd parity pseudoscalar curvature to the
Hilbert-Einstein-Cartan scalar curvature was first studied by Hojman,
Mukku, and Sayed \cite{Hojman:1980kv} (for Mukku's recent view see
\cite{Mukku:2007}):
\begin{eqnarray}\label{admixture}
  V_{\rm HMS}&=&\frac{1}{2\kappa}(a_0\,^{\star\!}R+b_0\,^{\star\!}X) \cr
  &=&-\frac{1}{2\kappa}(a_0\eta_{\a\b}
  +b_0\vt_{\a\b})\wedge R^{\a\b}\,.
\end{eqnarray}
Note that on the right-hand-side of this equation the star only enters
in $\eta_{\a\b}=\,^\star(\vt_\a\wedge\vt_\b)$, that
is, the second term $\vt_{\a\b}=\vt_\a\wedge\vt_\b$ is of odd parity
and thus $b_0$ is a pseudoscalar. The excitations turn out to be
\begin{equation}\label{admexcit}
H_\a=0\,,\quad H_{\a\b}=\frac{1}{2\kappa}(a_0\eta_{\a\b}+b_0\vt_{\a\b})\,.
\end{equation}
If we introduce the left-hand-side of (\ref{1stEC}) as the Einstein
3-form $G_\a:=\frac{1}{2}\eta_{\a\b\g}\wedge R^{\b\g}$, the field
equations read
\begin{eqnarray}
 a_{0}G_{\alpha} - b_{0}\,
^{\star}\! X_{\alpha} & = & \kappa{\Sigma}_{\alpha}\,,\label{EC_pv1}\\
& & \cr \frac{a_{0}}{2}\, T^{\mu}\wedge\,
{\eta}_{\alpha\beta\mu} + {b_{0}}\,
T_{[\alpha}\wedge\, {\vartheta}_{\beta ]} & = &\kappa
{\tau}_{\alpha\beta}\,.\label{EC_pv2}
\end{eqnarray}

The situation with respect to the second field equation (\ref{EC_pv2})
is similar as in the EC-theory. For vanishing material spin,
${\tau}_{\alpha\beta}=0$, the torsion vanishes, too, $T^\a=0$.  This
can be shown by substituting the irreducible decomposition of
torsion (\ref{irrtorsion}) into (\ref{EC_pv2}) and using the
geometrical identities
\begin{eqnarray}\label{eta_torsion}
^{(1)}T^{\mu}\wedge\, {\eta}_{\alpha\beta\mu} & = & 2\, ^{\star
(1)}T_{[\alpha}\wedge\, {\vartheta}_{\beta ]}\, ,\cr
^{(2)}T^{\mu}\wedge\, {\eta}_{\alpha\beta\mu} & = & -4\, ^{\star
(2)}T_{[\alpha}\wedge\, {\vartheta}_{\beta ]}\, ,\cr
^{(3)}T^{\mu}\wedge\, {\eta}_{\alpha\beta\mu} & = & -\, ^{\star
(3)}T_{[\alpha}\wedge\, {\vartheta}_{\beta ]}\, .
\end{eqnarray}
To supply non-vanishing torsion we either need material spin or at
least (for the vacuum case) field Lagrangians quadratic in the field
strengths. In this sense, the Lagrangian (\ref{admixture}) is as
degenerate as the one of EC and it is natural to turn to quadratic
odd parity Lagrangians. 

Subsequently questions related to the $V_{\rm HMS}$-Lagrangian, in the
realm of classical Riemann-Cartan spacetime, were investigated by
Nelson \cite{Nelson:1980ph}, Nieh and Yan \cite{Nieh:1981ww} (see also
Nieh's recent article\cite{Nieh:2007zz}), and McCrea et al.\
\cite{Bianchi,McCrea:1989sj}, see also Refs.\
\cite{Baekler:1992,Obukhov:1995eq}.

In the general context of the Ashtekar formalism or, more generally,
of {\it loop quantum gravity,} compare Kiefer \cite{Kiefer} and
Rovelli \cite{Rovelli}, the $V_{\rm HMS}$-Lagrangian was taken up
again by Holst \cite{Holst:1995pc}, Freidel et al.\
\cite{Freidel:2005sn,Freidel:2006hv}, Khriplovich et al.\
\cite{Khriplovich:2005jh}, and Bojowald et al.\
\cite{Bojowald:2007nu}; similar parity violating pieces were studied
by Mukhopadhyaya et al.\
\cite{Mukhopadhyaya:1998vp,Mukhopadhyaya:2002nu}, see also the related
papers by Mielke \cite{Mielke:2008zz,Mielke:2009}.

In the framework of a {\it quantum} field theoretical context,
Poplawski \cite{Poplawski:2010jv}, Randono \cite{Randono:2010ym}, and
Bjorken \cite{Bjorken:2010qx} developed cosmological models with
torsion and parity violating pieces that are induced by the vacuum
structure.

Jackiw and Pi \cite{JackiwPi} proposed a specific model with violation
of parity and Lorentz invariance in the context of GR. They
introduced, in addition to the Hilbert-Einstein Lagrangian, an {\it
  external} scalar field $\b$, not to be varied in the action
principle, multiplied by the Chern-Simons (CS) term attached to the
curvature:
\begin{equation}\label{LagrangianGRCS'}
  V_{\rm GRCS}=\frac{1}{2\kappa}\,^\star \widetilde{R}+
\frac{\beta}{2\varrho} \widetilde{R}_\a{}^\b\wedge
  \widetilde{R}_\b{}^\a\,.
\end{equation}
This model was extended to EC-theory by  Cantcheff \cite{Cantcheff},
\begin{equation}\label{LagrangianECCS'}
  V_{\rm ECCS}=\frac{1}{2\kappa}\,^\star R+\frac{\beta}{2\varrho} R_\a{}^\b\wedge
  R_\b{}^\a\,,
\end{equation}
see also Ertem \cite{Ertem}. Of course, both theories differ in their
physical content, as does GR from the EC-theory. We turn our attention to
the EC-version in (\ref{LagrangianECCS'}).

We know from geometry that a CS-term is an exact form. We have for a
RC-space, see \cite{PRs}, Eqs.(3.9.3) and (3.9.8),
\begin{equation}\label{boundary'}
-{1\over 2}R_\alpha{}^\beta\wedge
  R_\beta{}^\alpha=dC_{\rm RR}
\end{equation}
with
\begin{eqnarray}\hspace{-5pt}C_{\rm RR}&:=&
  -{1\over 2}(\Gamma_{\alpha}{}^{\beta}\wedge R_{\beta}
  {}^{\alpha} + {1\over 3}\Gamma_{\alpha}{}^{\beta}\wedge \Gamma_{\beta}
  {}^{\gamma}\wedge \Gamma_{\gamma}{}^{\alpha})\,.\label{CRR}
\end{eqnarray}
Thus,
\begin{equation}\label{LagrangianECCS''}
  V_{\rm ECCS}=\frac{1}{2\kappa}\,^\star R-\frac{\beta}{\varrho} dC_{\rm RR}\,.
\end{equation}
This Lagrangian contains an odd parity piece quadratic in curvature,
see (\ref{boundary'}). However, it is of a fairly degenerate
character. Still, since $\b$ is a prescribed field, the field
equations are affected by the CS-term. We will come back to an
explicit evaluation of the curvature square piece below.

Explicit odd parity curvature square pieces were present in some
cosmological models with spin $0^+$ and spin $0^-$ modes. These models
motivated us for a further search in this direction.

\subsection{Interlude: the impact of the cosmological model of
  Shie-Nester-Yo on PG}\label{Sec.IIIE}

Shie, Nester, and Yo \cite{Yo:2006qs,Shie:2008ms}, in the framework of
PG and in accordance with \cite{Yo:1999ex,Yo:2001sy}, formulated a new
cosmological model.  It contains, besides the graviton mode $2^+$ of
GR, one propagating connection and accordingly one propagating torsion
mode of spin $0^+$; the plus $+$ refers to the positive parity. The
corresponding Lagrangian reads effectively\footnote{Their actual
  Lagrangian \cite{Shie:2008ms}, Eq.~(18) contains also squared pieces
  of the axial torsion $^{(3)}T^\a$ and the tensor torsion
  $^{(1)}T^\a$. However, they find from the second field equation (for
  vanishing spin) that $^{(1)}T^\a={}^{(3)}T^\a=0$. Accordingly, in
  the SNY-model only the vector piece $^{(2)}T^\a$ of the torsion is
  active explicitly. An analogous remark applies to
  (\ref{extendedSNY}).}
\begin{equation} V_{\text{SNY}}=\label{SNY} \frac{1}{2\kappa}\left(a_0
    \,^\star R+\frac 13 a_{2} {\cal V}\wedge \,^\star  {\cal V}
  \right)-\frac{1}{24\varrho}w_6R^2\eta\,.
\end{equation}
This model has fairly realistic features and encouraged further
developments. Li et al.\ \cite{Li:2009zzc,Li:2009gj}, following Ref.\
\cite{Shie:2008ms}, investigated the cosmological evolution of the
SNY-model with advanced numerical techniques.

In order to embrace additionally connection modes with spin $0^-$, that is
with odd parity, in a next step, Chen et al.\cite{exSNY} generalized
the SNY-Lagrangian to
 \begin{eqnarray}\label{extendedSNY} V_{\text{SNY$'$}}&=&
   \frac{1}{2\kappa}\left(a_0 {}^\star R-2\lambda_0\eta\right)\nonumber\\
   &&+\frac{1}{6\kappa}\left(a_{2}  {\cal V}\wedge \,^\star  {\cal V}-
     a_{3}{ {\cal A}\wedge \,^\star {\cal A}}\right)
   \nonumber\\ &&
   -\frac{1}{24\varrho}\left(w_6R^2-w_3{X^2}\right)\eta\,.
\end{eqnarray}
It contains, as odd parity terms, the {\it axial vector torsion} $\cal
A$ and the {\it pseudoscalar curvature} $X$, see (\ref{axitor}) and
(\ref{irr3}), respectively. However, these odd parity terms are
concealed in an even parity Lagrangian since each of the terms $ {\cal
  A}\wedge \,^\star {\cal A}$ and ${X^2}\eta={X}\,^{\star\!} X$
contain one explicit star, respectively. Accordingly, this model with
propagating modes $2^+$ and $0^\pm$ is of even parity, but contains
concealed the odd parity terms $ {\cal A}$ and $X$.

It is then tempting to provide further add-ons, namely the parity
violating mixed terms $ {\cal A}\wedge{}^\star  {\cal V}$ and $X\,^\star R=R
\,^\star X$. This provides a further extension of the SNY-model with
respective metric and connection propagating modes of $2^{+}$ and
$0^{\pm}$.

\section{The shadow world of quadratic PG-Lagrangians with
odd parity terms}\label{Sec.IV}

\subsection{Constructing systematically quadratic odd parity
  Lagrangians}\label{Sec.IVA}

As discussed by Sozzi \cite{Sozzi}, {}from the validity of the CPT
symmetry in nature and {}from the fact that the C and CP symmetries are
only broken in weak interaction, but are valid otherwise, one would
expect roughly an equal amount of matter and antimatter in the
Universe. It appears however, as shown by Steigman
\cite{Steigman:1976ev}, that only what we call matter is around in the
Universe. Accordingly, we have to face the matter/antimatter asymmetry
in nature and may want to approach this question {}from a gravitational
point of view. Can we, in PG, construct odd parity Lagrangians in a
natural way so that we can estimate the possible influence of those
terms for the evolution of the Universe?\footnote{At the Large Hadron
  Collider (LHC) in Geneva the detector LHCb (the `b' = beauty refers
  to the `bottom' quark) was constructed particularly for
  investigations of matter-antimatter interactions, namely for the
  CP-violation in the interaction of b-hadrons. It is hoped that these
  experiments will provide new insight into the interrelationship between
  matter and antimatter.}

We proved identities of the type (\ref{identityw7}) under heavy use of
the computer algebra system Reduce, including the package Excalc for
handling directly exterior differential forms, see
\cite{Hearn,Schruefer,EXCALC,Socorro:1998hr}. In this way, we
can construct a quadratic odd parity Lagrangian $V_{-}$ that in its
general structure reflects the even parity Lagrangian $V_{+}$ in
(\ref{QMA}). It is the `shadow' of $V_{+}$:
\begin{eqnarray}\nonumber
 \label{parityodd}&&\hspace{-10pt}V_{-}=
-\frac{b_0}{2\kappa}\,^{(3)}R_{\alpha\beta}
  \wedge\vt^{\a\b} \cr & & \cr
& &+ \frac{1}{{\kappa}}\left( {\sigma}_{1}{}
^{(1)}T^{\alpha}\wedge\, ^{(1)}T_{\alpha} + {\sigma}_{2}{}
^{(2)}T^{\alpha}\wedge{} ^{(3)}T_{\alpha}\right)\cr & & \cr
  & &
-\frac{1}{2{\varrho}}\left( {\mu}_{1}{} ^{(1)}R^{\alpha\beta}\wedge{}
^{(1)}R_{\alpha\beta} + {\mu}_{2}{} ^{(2)}R^{\alpha\beta}\wedge{}
^{(4)}R_{\alpha\beta}\right. \\ & &
 + \left.
{\mu}_{3}{}^{(3)}R^{\alpha\beta}\wedge{} ^{(6)}R_{\alpha\beta} +
{\mu}_{4}{} ^{(5)}R^{\alpha\beta}\wedge{} ^{(5)}R_{\alpha\beta}
\right).
\end{eqnarray}
All its constants are pseudoscalars. In a Riemannian spacetime such an
extended shadow does not exists, since only the $\mu_1$-term survives,
because of $T^\a=0$ and
$^{(2)}R^{\alpha\beta}=\,^{(3)}R^{\alpha\beta}=\,^{(5)}R^{\alpha\beta}=0$. Therefore,
in PG we can free ourselves {}from the constraint to use even parity
Lagrangians with only one odd term; in fact, PG brings the existence
of numerous odd parity Lagrangians to light.

The special cases of Jackiw \& Pi \cite{JackiwPi} and Cantcheff
\cite{Cantcheff} can now be straightforwardly evaluated.  If we define
in four dimensions the (pseudo--)scalar product $\langle A,B\rangle$
for any two tensor--valued two--forms $A_{\a_1\ldots\a_r}$ and
$B_{\a_1\ldots\a_r}$ by
\begin{equation}\langle A,B\rangle := \,^\star(A_{\a_1\ldots\a_r}\wedge 
B^{\a_1\ldots\a_r})\,,
\end{equation}
we can write the curvature square piece of our odd Lagrangian
(\ref{parityodd}) as
\begin{eqnarray} \label{parityodd2'}\nonumber
  \stackrel{{\text{curv}^2}}{V_{-}}& =&
  \frac{1}{2{\varrho}}\,^\star\!\left( {\mu}_{1}{}\langle
    ^{(1)}R,{}^{(1)}R\rangle + {\mu}_{2}{}\langle
    ^{(2)}R,{}^{(4)}R\rangle\right.\\ &&\left. + {\mu}_{3}{}\langle
    ^{(3)}R,{}^{(6)}R\rangle + {\mu}_{4}{}\langle
    ^{(5)}R,{}^{(5)}R\rangle\right).
\end{eqnarray}

On the other hand, the ECCS-Lagrangian reads
\begin{equation}\label{LagrangianECCS1'}
  V_{\rm ECCS}=\frac{1}{2\kappa}\,^\star
  R+\frac{\beta}{2\varrho}\,^{\star} \langle R,R\rangle \,.
\end{equation}
If we substitute the irreducible pieces of the curvature into the
scalar product, we find [see \cite{PRs}, Eq.(B.4.37)]
\begin{eqnarray}\label{curvsquare'}\nonumber
  \langle R,R\rangle &=& \langle\,^{(1)}\!R,\,^{(1)}R\rangle
  +2\,\langle\,^{(2)}\!R,\,^{(4)}R\rangle\\ &&
  +2\,\langle\,^{(3)}\!R,\,^{(6)}R\rangle+\langle\,^{(5)}\!R,\,^{(5)}R
  \rangle\,.
\end{eqnarray}
In a Riemannian space of the theory of Jackiw \& Pi only the
conformally invariant Weyl square piece, the first term on the
right-hand-side of (\ref{curvsquare'}), is left over, whereas the
RC-space complicates the structures by additional post-Riemannian
pieces. Comparing (\ref{parityodd2'}) with (\ref{curvsquare'}), we
find
\begin{equation}
  \stackrel{{\text{curv}^2}}{V_{-}}\!\!\left(\mu_1=1,
    \mu_2=2,\mu_3=2,\mu_4=1
  \right)=-\frac{1}{\varrho}dC_{\rm RR}\,.
\end{equation}
In other words, our odd parity curvature square Lagrangians, for the
coupling constants specified, becomes an exact form. Consequently only
three of the four $\mu$'s can be chosen independently.

The mixed quadratic Lagrangian with even and odd parity is then
\begin{equation}\label{LagrangianPM}
V_{\pm}=V_{+}+V_{-}\,,
\end{equation}
see also \cite{YuriEtAl}.

\subsection{Cosmological model with parity violating terms}\label{Sec.IVB}

If we compare $V_{\pm}$ with $V_{\text{SNY$'$}}$, the next
step of `minimally' generalizing (\ref{extendedSNY}) and hopefully
keeping the nice properties of the model is to allow for unconcealed
odd parity pieces, but only those odd parity pieces that already occur
in (\ref{extendedSNY}), namely $X$ and ${\cal A}$. Thus, starting with
(\ref{LagrangianPM}) and putting the following constants to zero,
\begin{align}\label{zeroconstants}
  a_1=0,\sigma_1= 0; w_1={}&w_2=w_4=w_5=0;\nonumber\\
  \mu_1={}&\mu_2=\mu_4=0,
\end{align}
we arrive at the new Lagrangian
\begin{align}
  &V_{\rm BHN} =
  \frac{1}{2\kappa}\!\left(\!\!-a_0{}^{(6)}R^{\alpha\beta}\wedge\eta_{\alpha\beta}
    -b_0{}^{(3)}R^{\alpha\beta}\wedge\vt_{\a\b}
    -2\lambda_{0}\eta \right.\nonumber \\
  &\left.\hspace{55pt}+ a_2{}^{(2)}T^\a\wedge{}^{\star(2)}T_\a
    +a_3{}^{(3)}T^\a\wedge{}^{\star(3)}T_\a\right. \nonumber \\
  &\left.\hspace{55pt} +2\sigma_2{}^{(2)}T^\a\wedge{}^{(3)}T_\a\right)
  \nonumber \\ &\hspace{10pt}-\frac{1}{2\varrho}
  \left(w_3{}^{(3)}R^{\alpha\beta} \wedge{}^{\star(3)}
    R_{\alpha\beta}+ w_6{}^{(6)}R^{\alpha\beta} \wedge{}^{\star(6)}
    R_{\alpha\beta}\right. \nonumber \\ &\left.\hspace{55pt} +
    \mu_3{}^{(3)}R^{\alpha\beta} \wedge{}^{(6)} R_{\alpha\beta}
  \right)\,.\label{eben}
\end{align}
Substituting the irreducible pieces of the torsion (\ref{trator}) and
(\ref{axitor}) and of the curvature (\ref{irr3}) and (\ref{W6}) into
(\ref{eben}), we find the {\it more compact form of our new gravitational
Lagrangian}
\begin{eqnarray}\label{L_BHN'}
  V_{\rm BHN} & = &
  \frac{1}{2{\kappa}}(a_{0}{}^{\star\!} R+b_{0}{}^{\star\!\!}{X} -
  2{\lambda}_{0}{\eta})\nonumber\\
  &&\hspace{-7pt}+\frac{1}{6{\kappa}}\left(a_{2} {\cal V}\wedge
    {}^{\star\!} {\cal V} -
    a_{3}{{\cal A}\wedge{} ^{\star\!\!\!}{\cal A}}
    -2{\sigma}_{2}{ {\cal V}\wedge{}
      ^{\star\!\!\!} {\cal A}}\right)\nonumber\\
  &&\hspace{-7pt}-\frac{1}{24{\varrho}}\left( w_{6}R\,^{\star\!}R- w_{3}
    {X\,^{\star\!\!} X} +
    {\mu}_{3}R\,^{\star\!\!} X\right)\,.
\end{eqnarray}
The constants $b_0,\,\sigma_2,\,\mu_3$ are {\it pseudo\/}scalar, the
remaining ones are scalar.\footnote{A teleparallel Lagrangian with a
  term with $\sigma_2\ne 0$ has been considered earlier by
  M\"uller-Hoissen and Nitsch \cite{MHN83}.}

We can read off from the Lagrangian a symmetry between $R$ and $X$,
between $w_6$ and $w_3$, and between $a_0$ and $b_0$. There is also a
symmetry between $ {\cal V}$ and $\cal A$ on the one side and $R$ and $X$ on
the other side; this implies that $a_2,a_3,\sigma_2$ are mirrored in
$w_6,w_3,\mu_3$. These symmetries are also reflected in the field
equations. We will come back to this in Sec.~\ref{Sec.IVE}.

The decomposition of the linear terms in $R$ and $X$ into Riemannian
and post-Riemannian pieces, modulo surface terms, yields
\begin{eqnarray}\label{BHNdecomp}
  V_{\rm BHN}&=&\frac{1}{2\kappa}\left[a_0\,^\star\widetilde{R}
    -2{\lambda}_{0}{\eta}
    +\,^{(1)}T_\a\wedge\left(b_0 + a_0\,^\star
    \right)^{(1)}T^\a \right.\nonumber \\
  &&\left.-\frac 23 m^+ {\cal V}\wedge\,^\star  {\cal V}+\frac 16 m^-{\cal
      A}\wedge\,^{\star\!\!}{\cal A} - \frac 23 m^\times {\cal
      V}\wedge\,
^{\star\!\!} {\cal A}\right]\nonumber\\
    &&-\frac{1}{24{\varrho}}\left( w_{6}R\,^{\star\!}R- w_{3}
      {X\,^{\star\!\!} X} +
      {\mu}_{3}R\,^{\star\!\!} X\right)\,.
\end{eqnarray}
with
\begin{equation}\label{emms}m^+:=a_0-\frac{a_2}{2}\,,\;
  m^-:=a_0-2{a_3}\,,\;
m^\times:=b_0+\sigma_2\,.
\end{equation}
We shall see that the $m$'s play a role in the discussion of the
second field equation. Note that $m^+$ and $m^-$ are of even parity,
whereas $m^\times$ is odd. A corresponding decomposition of the
quadratic curvature terms in (\ref{BHNdecomp}) doesn't seem to provide
new insight.

\subsection{Diagonalization of the BHN-Lagrangian}\label{Sec.IVC}

\subsubsection{Eigenvalues of the kinetic matrix of the translational
  gauge potential $\vt^\a$}\label{Sec.IVC1}

In the Lagrangian (\ref{BHNdecomp}) the term ${\cal V}\wedge \,
^{\star} {\cal A}={\cal A}\wedge \,
^{\star} {\cal V}={\cal A}_{\mu} {\cal V}^{\mu}{\eta} $ represents an
interaction term of two four-vectors of the type {\em vector $\times$
  axial vector}. To get some more insight into the dynamics of the
fields governed by the Lagrangian (\ref{BHNdecomp}), we decompose
these four-vectors into $(1\oplus 3)$ with
\begin{equation} {\cal A}_{\mu}=\left({\cal A}_{0}\, , {\vec{\cal
        A}}\,\right)\quad {\rm and}\quad  {\cal V}_{\mu}=
\left( {\cal V}_{0}\, , {\vec     {\cal V}}\, \right)\,.
\end{equation}
The translational part of the Lagrangian (\ref{BHNdecomp}), with
$^{(1)}T^{\alpha}=0$, is proportional to the quadratic form
\begin{eqnarray}\label{q_form}
{\cal Q} &:=& 4( {\cal V}_{0}^2- {\cal V}_{1}^2
- {\cal V}_{2}^2- {\cal V}_{3}^2)m^{+}\nonumber\\ & &\hspace{-3pt}- ({\cal
  A}_{0}^2-{\cal A}_{1}^2-{\cal A}_{2}^2- {\cal
  A}_{3}^2)m^{-}\nonumber\\ & &\hspace{-8pt}+4({\cal A}_{0} {\cal V}_{0}
-{\cal A}_{1} {\cal V}_{1}-{\cal A}_{2} {\cal V}_{2}- {\cal A}_{3}
 {\cal V}_{3})m^{\times}\!.
\end{eqnarray}
Expressed in terms of matrices, we have
\begin{eqnarray}\label{matrixeq}
{\cal Q} & = & ( {\cal V}_{0}\, ,{\vec  {\cal V}}\, , {\cal A}_{0}\, ,
 {\vec{\cal A}}\,)\cdot \left(
\begin{matrix}-4m^{+}{\bf g} & -2m^{\times}{\bf g}\cr
-2m^{\times}{\bf g} & m^{-}{\bf g}\end{matrix}\right)\!\cdot\!
\left( \begin{matrix}  {\cal V}_{0}\cr {\vec  {\cal V}}\cr {\cal A}_{0}\cr
{\vec{\cal A}}\end{matrix}\right)\! ,\cr & = & ( {\cal V}_{0}\, ,{\vec
 {\cal V}}\, , {\cal A}_{0}\, , {\vec{\cal A}}\,)\cdot {\cal T}\cdot
( {\cal V}_{0}\, ,{\vec  {\cal V}}\, , {\cal A}_{0}\, 
, {\vec{\cal A}}\,)^{\text{T}}\,,
\end{eqnarray}
with {\bf g} as the four-dimensional Minkowski metric.

With the useful abbreviations
\begin{equation}x:=4m^{+}\, , \quad y:=2m^{\times}\, , \quad
  z:=m^{-}\,,
\end{equation}
the new matrix ${\cal T}$ reads
\begin{eqnarray}\label{cal_T}
{\cal T} & = & \left( \begin{matrix} x & 0 & 0 & 0 & y & 0 & 0 &
0\cr 0 & -x & 0 & 0 & 0 & -y & 0 & 0 \cr 0 & 0 & -x & 0 & 0 & 0 &
-y & 0 \cr 0 & 0 & 0 & -x & 0 & 0 & 0 & -y \cr y & 0 & 0 & 0 & -z
& 0 & 0 & 0 \cr 0 & -y & 0 & 0 & 0 & z & 0 & 0 \cr 0 & & -y & 0 &
0 & 0 & z & 0 \cr 0 & 0 & 0 & -y & 0 & 0 & 0 & z \end{matrix}
\right)\cr & = & \left( \begin{matrix} -x{\bf g} & -y{\bf g}\cr
-y{\bf g} & z{\bf g}\end{matrix}\right).
\end{eqnarray}
It has some simple properties:
\begin{eqnarray}
{\cal T}^{\text{T}} & = & {\cal T}\,,\cr {\rm trace}\,{\cal T} & = &
  -2(x-z)\,,\cr {\rm det}\,{\cal
    T} & = & \left(xz+y^2\right)^4 > 0\,,\cr {\cal T}^{-1} & = &
    \frac{1}{xz+y^2}\left( \begin{matrix} -z{\bf g} & -y{\bf g}\cr
-y{\bf g} & x{\bf g}\end{matrix}\right)\, .
\end{eqnarray}

In the following we will make use of Dirac's bra-ket notation, see
Dirac \cite{Dirac:1930} and Schouten \cite{Schouten}. For this purposes
we define (abstract) vectors according to
\begin{equation}\label{kets}
<{\cal X}| := ({\cal V}_{0}\, ,{\vec {\cal V}}\, , {\cal A}_{0}\,, {\vec{\cal
A}}\,)\; {\rm and}\;|{\cal X}> :=({\cal V}_{0}\, ,{\vec {\cal V}}\, ,
{\cal A}_{0}\,, {\vec{\cal A}}\,)^{\text{T}}
\end{equation}
such that the quadratic form ${\cal Q}$ becomes
\begin{equation}\label{calQ2}
{\cal Q} = <{\cal X}| {\cal T} |{\cal X}>\,.
\end{equation}
To diagonalize the form (\ref{calQ2}), we introduce a new vector
$|{\cal Y}>$ together with a suitable  orthonormal matrix ${\cal
K}$ such that
\begin{equation}\label{calX2}
|{\cal X}> =: {\cal K}|{\cal Y}> \quad {\rm and}\quad <{\cal X}|
= <{\cal Y}|{\cal K}^{\text{T}}\,.
\end{equation}
Substitution of (\ref{calX2}) into (\ref{calQ2}) yields the
covariant expression
\begin{eqnarray}
{\cal Q} & = & <{\cal X}|{\cal T}|{\cal X}> = <{\cal Y}|{\cal
K}^{\text{T}}\cdot {\cal T}\cdot {\cal K}|{\cal Y}> \cr & = & <{\cal
Y}|{\cal D}|{\cal Y}>\,.
\end{eqnarray}
We will choose the orthonormal matrix ${\cal K}$ such that the product
${\cal D}:={\cal K}^{\text{T}}\cdot {\cal T}\cdot {\cal K}$ is diagonal and
thus contains the eigenvalues of
${\cal T}$ as entries.

The eigenvalues of $\cal T$ turn out to be:
\begin{eqnarray}\label{eigenvalues_T}
  {\Lambda}_{1} & = & \frac{1}{2}\left(x-z +
    \sqrt{(x+z)^2+4y^{2}}\right),\cr & & \cr
  {\Lambda}_{2} & = & \Lambda_{3}
  = {\Lambda}_{4}  = -{\Lambda}_{1}\,,\cr & & \cr
  {\Lambda}_{5} & = & \frac{1}{2}\left(x-z -
    \sqrt{(x+z)^2+4y^{2}}\right),\cr & & \cr
  {\Lambda}_{6} & = &
  \Lambda_{7} = {\Lambda}_{8}= -{\Lambda}_{5}\,.
\end{eqnarray}
According to the Lorentz structure, that is, the $(1\oplus
3)$-decomposition of four-vectors, we have two different
eigenvalues for the time components and 2 threefold eigenvalues
for the spatial components, respectively.

For the explicit construction of the matrix ${\cal K}$ we need the
eigenvectors $\vec{u}_{\Lambda_n}$ of ${\cal T}$, where $n$ runs
from 1 to 8. Their components can be expressed (for $y\neq 0$) in
terms of $A$ and $B$:
\begin{eqnarray}
  \hspace{-10pt}  A  \hspace{-5pt}& := & \hspace{-5pt}
  -\frac{1}{2y}\left(x+z+\sqrt{(x+z)^2
      +4y^2}\right),\cr \hspace{-10pt} & &
  \cr B \hspace{-5pt} & := & \hspace{-5pt}
  - \frac{1}{2y}\left(x+z-\sqrt{(x+z)^2+4y^2}\right).
\end{eqnarray}
We normalize the eigenvectors by means of
\begin{equation}\label{norm}
 a  :=  \sqrt{1+A^2}\qquad\text{and}\qquad b := \sqrt{1+B^2}\,.
\end{equation}
The columns of the matrix ${\cal K}$ are the normalized
eigenvectors of the matrix ${\cal T}$. Accordingly, we find $\cal
K=$
\begin{equation}\label{cal_K}
\left( \begin{matrix} -A/a & 0 & 0 & 0 & -B/b & 0 & 0 & 0\cr 0 &
1/b & 0 & 0 & 0 & -1/a & 0 & 0\cr 0 & 0 & -A/a & 0 & 0 & 0 & -B/b
& 0\cr 0 & 0 & 0 & 1/b & 0 & 0 & 0 & -1/a \cr 1/a & 0 & 0 & 0 &
1/b & 0 & 0 & 0\cr 0 & B/b & 0 & 0 & 0 & -A/a & 0 & 0\cr 0 & 0 &
1/a & 0 & 0  & 0 & 1/b & 0\cr 0 & 0 & 0 & B/b & 0 & 0 & 0 & -A/a
\end{matrix}\right).
\end{equation}
Because of the identity $BA=-1$ and its consequences $-A/a=1/b$ and
$B/b=1/a$, there is actually much more symmetry than is apparent in
the matrix (\ref{cal_K}), which in fact depends essentially on only one
parameter.  This matrix has simple properties like ${\rm det}\,{\cal
  K}=+1$, ${\cal K}\cdot {\cal K}^{\text{T}}={\bf 1}_{8\times 8}$; the
eigenvalues are $e^{\pm i\alpha}$, both with multiplicity four and
with $\tan\alpha=1/A$, cf.\ Eq.\ (\ref{cs}). Hence in the
eight-dimensional vector
space the matrix ${\cal K}$ represents a pure rotational matrix.

Thus, the quadratic form ${\cal Q}$ assumes the diagonal form
\begin{eqnarray}\label{matrix2}
  &&\hspace{-30pt}{\cal Q}  =  <{\cal X}|{\cal T}|{\cal X}>\cr &&
  \hspace{-30pt}= <{\cal Y}|
  {\rm diag}({\Lambda}_{1}\, ,-{\vec{\Lambda}_{1}}\, , {\Lambda}_{5}\,
  ,   -{\vec{\Lambda}}_{5})|{\cal Y}>\,,
\end{eqnarray}
with the obvious abbreviations
${\vec{\Lambda}_{1}}:={\Lambda}_{1}(1,1,1)$ and
${\vec{\Lambda}_{5}}:={\Lambda}_{5}(1,1,1)$. The new vector
$|{\cal Y}>$ turns out to be
\begin{equation}\label{vector}
|{\cal Y}>:=\left(
\begin{matrix} {\bf {V}}_{0}\cr {{ \vec {\bf {V}}}}\cr
 { {\cal {\bf A}}}_{0}\cr
{{ \vec{\bf A}}}\end{matrix}\right)={\cal K}^{\text{T}}|{\cal X}>={\cal
K}^{\text{T}}\cdot \left(
\begin{matrix} {\cal V}_{0}\cr {\vec {\cal V}}\cr {\cal A}_{0}\cr
{\vec{\cal A}}\end{matrix}\right).
\end{equation}
Accordingly, the quadratic form (\ref{q_form}) can now be written in
diagonal form as
\begin{equation}\label{quadrat}
{\cal Q}={\Lambda}_{1}\left( {{\bf {V}}_{0}}^2 - {{\vec {\bf
{V}}}}^2\right) + {\Lambda}_{5}\left( {{\bf A}_{0}}^2 - {{\vec{\bf
A}}}^2\right)\,.
\end{equation}

We need to pay special attention also to the case $y=0$, namely when
there is no coupling between the vector and the axial vector of the
torsion. In this case, the matrix ${\cal T}$ is diagonal with
\begin{equation}\label{calT_y0}
{\cal T}(y=0) = \left( \begin{matrix} -x{\bf g} & 0\cr 0 & z{\bf
g}\end{matrix} \right)\,,
\end{equation}
and the eigenvectors of (\ref{calT_y0}) correspond to the
eight-dimensional unit vector. Thus, for $y\rightarrow 0$, the
transformation matrix becomes the unit matrix: ${\cal K}={\bf
  1}_{8\times 8}$. The transformation to the diagonal matrix yields
\begin{equation}
\hspace{-2pt}{\cal K}^{\text{T}}{\cal T}{\cal K} = {\cal D}={\rm
diag}(x,-x,-x,-x,-z,z,z,z).
\end{equation}
In this case, the quadratic form ${\cal Q}$ reduces to
\begin{eqnarray}
  {\cal Q}(y\rightarrow 0) & = & x\left( {{\bf {V}}_{0}}^2 -
    {{\vec {\bf {V}}}}^2\right) - z\left( {{\bf A}_{0}}^2 -
    {{\vec{\bf A}}}^2\right)\,\cr 
  = 4m^{+}&&\hspace{-20pt}\left( {{\bf {V}}_{0}}^2 - {{\vec {\bf
          {V}}}}^2\right) - m^{-}\left( {{\bf A}_{0}}^2 -
    {{\vec{\bf A}}}^2\right).
\end{eqnarray}

\subsubsection{Representation of $\cal K$ in terms of angles}\label{Sec.IVC2}

The matrix ${\cal K}$ as a rotation matrix can be parametrized by
introducing a suitable angle variable.  For this purpose we use
\begin{equation}\label{cs}
\sin\alpha =\frac{1}{\sqrt{1+A^2}}\quad\text{and}\quad \cos\alpha=
\frac{A}{\sqrt{1+A^2}}\,.
\end{equation}
In terms of this parameter ${\alpha}$, the matrix ${\cal K}$ can be
displayed in compact form as
\begin{equation}
{\cal K}=\left(
          \begin{array}{cc}
            \>\cos\a\>{\bf I_{4\times 4}} & \sin\a\>{\bf I_{4\times 4}} \\
            -\sin\a\>{\bf I_{4\times 4}} & \cos\a\>{\bf I_{4\times 4}} \\
          \end{array}
        \right),\label{newK}
\end{equation}
where ${\bf I_{4\times 4}}$ is the $4\times4$ unit matrix
and $(x,y,z)$ are related to the parameter ${\alpha}$ by
\begin{equation}
\tan\a=\frac{x+z-\beta}{2y}=\frac{-2y}{x+z+\beta}\,,\label{angledef}
\end{equation}
with
\begin{equation}
\beta:=\sqrt{(x+z)^2+4y^2}.\end{equation}
Thus, the matrix product ${\cal T}\cdot{\cal K}$ reduces to
\begin{equation}
{\cal T}\cdot{\cal K} = \left(
          \begin{array}{cc}
             \> \cos\a\>{\bf I_{4\times 4}} & \sin\a\>{\bf I_{4\times 4}} \\
            -\sin\a\>{\bf I_{4\times 4}} & \cos\a\>{\bf I_{4\times 4}} \\
          \end{array}
        \right)\cdot {\cal D} = {\cal K}\cdot {\cal D}\, ,
\end{equation}
the condition for a similarity transformation.

\subsubsection{A possible parameter set}\label{Sec.IVC3}

We can read off from (\ref{eigenvalues_T}) the relations
${\Lambda}_{1}{\Lambda}_{2}<0$ and ${\Lambda}_{5}{\Lambda}_{6}<0$. As
an example, let us consider the case ${\Lambda}_{1}>0$ and
${\Lambda}_{5}>0$. With these assumptions we can define
\begin{eqnarray}\label{tau_alpha}
  \hspace{-18pt}{\nu}_{0}\hspace{-5pt} & := &\hspace{-5pt}
  {\sqrt{\Lambda}_{1}}{\bf V_{0}}\, ,\quad\hspace{7pt}
  {\vec{\nu}}:={\sqrt{\Lambda}_{1}}\,{\vec{\bf V}}\, , \quad
  {\Lambda}_{1}>0,\cr\hspace{-18pt} {\alpha}_{0}
  \hspace{-5pt}  &
  := &\hspace{-5pt}  {\sqrt{\Lambda}_{5}}{\bf A_{0}}\, ,\quad\hspace{7pt}
  {\vec{\alpha}}:={\sqrt{\Lambda}_{5}}\, {\vec{\bf A}}\, ,
  \quad {\Lambda}_{5}>0.
\end{eqnarray}
With (\ref{tau_alpha}), the quadratic form ${\cal Q}$
(\ref{quadrat}) assumes its `special relativistic' appearance
\begin{eqnarray}\label{Q_rel}
{\cal Q} & = & \left({\nu}_{0}^2 - {\vec{\nu}}\, ^2 \right) +
\left({\alpha}_{0}^2 - {\vec{\alpha}}\, ^2 \right)\cr & = &
-g_{\mu\nu}\left(
{\nu}^{\mu}{\nu}^{\nu}+{\alpha}^{\mu}{\alpha}^{\nu}\right)\,.
\end{eqnarray}
This is the sum of the squares of two four-vectors in a suitable
orthonormal reference frame and the Lorentz covariance is manifest.

Let us analyze the conditions to be imposed provided one assumes
$\Lambda_1>0,\Lambda_5>0$. From (\ref{eigenvalues_T}) we derive the
constraints
\begin{eqnarray}\label{range_I1}
x-z>0 & \quad\Longleftrightarrow\quad & 4m^{+}-m^{-}>0 \cr &\Longleftrightarrow&
3a_{0}-2(a_{2}-a_{3})>0\, ,
\end{eqnarray}
and
\begin{eqnarray}\label{range_I2}
  &&xz+y^2<0 \quad\Longleftrightarrow\quad  m^{+}m^{-}+(m^{\times})^2 < 0
 \quad \Longleftrightarrow\quad\cr
  &&\hspace{10pt}\left(
    a_{0}-\frac{a_{2}}{2}\right)(a_{0}-2a_{3})+(b_{0}+{\sigma}_{2})^2
  < 0\,.
\end{eqnarray}
On the other hand, assuming instead $\Lambda_1>0$, $\Lambda_5<0$ leads
to the condition
\begin{eqnarray}\label{condition}
&&xz+y^2>0\quad\Longleftrightarrow\quad m^+m^-+(m^\times)^2>0\quad
\Longleftrightarrow\quad\cr
&&\hspace{10pt}\left(
    a_{0}-\frac{a_{2}}{2}\right)(a_{0}-2a_{3})+(b_{0}+{\sigma}_{2})^2
  > 0\,.
\end{eqnarray}
One could similarly find the parameter conditions associated with the other two cases.

If ${\bf V}_\mu$ and ${\bf A}_\mu$ are both timelike---as they will
turn out to be for the cosmological model which we derive
below---every set of parameters fulfilling the inequalities
(\ref{range_I1}) and (\ref{range_I2}) will lead to a strictly positive
kinetic energy matrix for the translational gauge potentials.


\subsubsection{Eigenvalues of the kinetic matrix of the Lorentz gauge
  potential $\Gamma^{\a\b}$}\label{Sec.IVC4}

Similar as in Sec.~\ref{Sec.IVC1}, we consider the quadratic form ${\cal C}$
representing the curvature square terms in (\ref{BHNdecomp}) and hence
the kinetic parts of the connection. This quadratic form is given by
\begin{eqnarray}
{\cal C} & := & w_{6}R^2-w_{3}X^2+\mu_{3}RX\cr & = &(R,X)\cdot
\left(\begin{matrix} w_{6} & {\mu}_{3}/2\cr {\mu}_{3}/2 & -
w_{3}\end{matrix}\right)\cdot \left(\begin{matrix}R\cr
X\end{matrix}\right)\cr & = &
 <{\cal Z}|{\cal B}|{\cal Z}>\,.
\end{eqnarray}
We wish to diagonalize the {\em symmetric} $(2\times2)$-matrix
\begin{equation}
{\cal B}:=\left(
  \begin{array}{cc}
    w_6 & \mu_3/2 \\
    \mu_3/2 & -w_3 \\
  \end{array}
\right).\label{curvmatrix}
\end{equation}
whose eigenvalues are
 \begin{equation}
\lambda_{1,2}=\frac12( w_6-w_3\pm\sqrt{\chi}),\quad
   \chi:=(w_6-w_3)^2+\Delta,
 \end{equation}
 where
 \begin{equation}\label{BBB} \Delta:=-4\det{\cal B}=4w_3w_6+\mu_3^2.
\end{equation}
This diagonalization can be simply accomplished with a rotation
matrix whose columns are orthogonal unit eigenvectors; the
transformation matrix and the eigenvectors can be parameterized by
a single angle $\gamma$ that can be determined from
\begin{equation}
\tan
\gamma
=\frac{w_6+w_3-\sqrt{\chi}}{\mu_3}=\frac{-\mu_3}{w_6+w_3+\sqrt{\chi}}.
\end{equation}
Then,
\begin{equation}
{\cal M}^{\text{T}}{\cal B}{\cal M}=\left(
               \begin{array}{cc}
                 \lambda_1 & 0 \\
                 0 & \lambda_2 \\
               \end{array}
             \right)\!,\,
{\cal M}:=\left(
\begin{array}{cc}
      \cos\gamma & \sin\gamma \\
      -\sin\gamma & \cos\gamma \\
    \end{array}
  \right).
\end{equation}
Consequently, with
\begin{equation}
({\widehat R},{\widehat X})=(R,X)\cdot{\cal M},
\end{equation}
the quadratic form ${\cal C}$ can be expressed as
\begin{eqnarray}
{\cal C} & = & < {\cal Z}|{\cal B}|{\cal Z} > = <{\cal Z}{\cal
M}|{\cal M}^{\text{T}}{\cal B}{\cal M}|{\cal M}^{\text{T}}{\cal Z}>\cr & = &
<{\widehat{\cal Z}}|{\cal D}_{R}|{\widehat{\cal Z}}> =
{\lambda}_{1}{\widehat R}^2 + {\lambda}_{2}{\widehat X}^2\,,
\end{eqnarray}
where ${\cal D}_{R}:={\cal
M}^{\text{T}}{\cal B}{\cal M}$ denotes a diagonal matrix.

The quadratic curvature terms in the Lagrangian can now be rewritten in
the form
\begin{eqnarray}\label{eq9:120}
V_{{\rm R}^2}  =  -\frac{1}{24\varrho}\left(
{\lambda}_{1}{\widehat R}^2+{\lambda}_{2}{\widehat
X}^2\right){\eta}\,.
\end{eqnarray}

As an illustrative example, consider in particular the case where
${\lambda_1}$ and $\lambda_2$ are both negative. This
  immediately leads to the following constraints on the coupling
  constants:
\begin{equation}\label{con_1}
w_{6}-w_{3}<0\, , \quad {\rm and}\quad {\mu}_{3}^2+4w_{3}w_{6}<0\,
,
\end{equation}
from which we can infer that
\begin{equation}
w_{6}<0,\quad {\rm and}\quad w_{3}>0
\end{equation}
for the aforementioned case, ${\lambda}_{1}<0$ and
${\lambda}_{2}<0$. Then we can rescale the variables and introduce
new ones according to
\begin{equation}
{\bf R}:=\frac{{\widehat R}}{\sqrt{{|\lambda}_{1}|}}\quad {\rm
and} \quad {\bf X}:=\frac{{\widehat X}}{\sqrt{{|\lambda}_{2}|}}
\end{equation}
such that the quadratic form ${\cal C}$ in this particular case
becomes
\begin{equation}
{\cal C}=-\left({\bf R}^2 + {\bf X}^2\right)\,.
\end{equation}
For the three other cases, namely $(\lambda_1>0,\lambda_2>0)$,
$(\lambda_1<0,\lambda_2>0)$, $(\lambda_1>0,\lambda_2<0)$, one can
do an analogous rescaling.


\subsubsection{Partly diagonalized Lagrangian}\label{Sec.IVC5}

Thus, the process of diagonalization for the case
$\Lambda_1>0,\Lambda_5>0,\lambda_1<0,\lambda_2<0$ leads to the
following diagonal pieces of the $V_{\rm BHN}$-Lagrangian,
\begin{eqnarray}
V_{\text{T$^{2}$}} & = & \frac{1}{12\kappa}\left[
\left({\nu}_{0}^2 - {\vec{\nu}}\,^2\right) + \left(
{\alpha}_{0}^2 - {\vec{\alpha}}\,^2\right)\right]{\eta} =
\frac{1}{12\kappa}\, {\cal Q}{\eta}\,,\cr & & \cr
V_{\text{R$^{2}$}} & = & \frac{1}{24{\varrho}}\left( {\bf R}^2 +
{\bf X}^2\right){\eta} = -\frac{1}{24\varrho}\,{\cal C}{\eta}\,
\end{eqnarray}
(with analogous results for the other sign choice cases).
Collecting the results received so far, we can give a new
representation of the Lagrangian (\ref{BHNdecomp}) in the form of
\begin{eqnarray}\label{V_diagonal}
V_{\rm BHN}  = &
&\hspace{-10pt}\frac{1}{2\kappa}\left(a_{0}{\widetilde
R}-2{\lambda}_{0}\right){\eta} \cr & &\hspace{-13pt}
+\frac{1}{2{\kappa}}\left(a_{0}\, ^{\star(1)}T^{\alpha}\wedge\,
^{(1)}T_{\alpha} + b_{0}\, ^{(1)}T^{\alpha}\wedge\,
^{(1)}T_{\alpha}\right) \cr & &\hspace{-13pt}
+\frac{1}{12\kappa}\, {\cal Q}{\eta}-\frac{1}{24\varrho}\,{\cal
C}{\eta}\,.
\end{eqnarray}

\subsubsection{Correspondences of eigenvalues of the kinetic matrices to
  spin $\&$ parity}\label{Sec.IVC6}

In this section we considered sufficient conditions for the
coefficients of the kinetic energy matrix being positive. We now
assume that the trace and the axial-vector pieces of the torsion are
both propagating independently. Then we require that the four-vectors
${\nu}^{\mu}$ and ${\alpha}^{\mu}$ are timelike during the whole
evolution. Accordingly, the propagation of independent {\it massive}
modes is characterized by ${\nu}^{\mu}{\nu}_{\mu}<0$ {and}
${\alpha}^{\mu}{\alpha}_{\mu}<0$. This is met by the requirement
(\ref{tau_alpha}). Other choices of the signs of ${{\Lambda}_{k}}$'s
will lead to spacelike four-vectors. The null case will be treated
separately in a continuation of this paper.

The Lagrangian (\ref{V_diagonal}) admits the introduction of a number
of strictly positive functions, that is, functions of
${{\Lambda}_{k}}$'s, such that for each of its dynamical variables we
can associate to each eigenvalue of the kinetic energy matrix the
corresponding spin \& parity state.

If we decompose the four-dimensional one-forms $\cal V$ and ${\cal A}$ into
$(1\oplus 3)$, respectively, we are naturally led to their spin
contents. Namely, we can introduce (massive) three-dimensional vectors
${\vec {\nu}}$ and ${\vec {\alpha}}$ and can associate to each of
them a corresponding three-dimensional spin \& parity
state. This method is not sensitive to a possible occurrence of
multiplicities of spin \& parity states.  For this purposes, we have to
investigate the corresponding lower-dimensional subspaces.

In our model, with $^{(1)}T^{\alpha}=0$, we have only propagating {\it
  scalar} and three-dimensional {\it vector} modes.  In a heuristic
manner, the diagonalization allows for the following tentative
correspondences:
\begin{eqnarray}
  \Lambda_1({\bf V}_{0}) & \lapl & \, 0^{+}\cr \Lambda_1({\vec {\bf
      V}}\,)
  & \lapl \,& 1^{+}\cr \Lambda_5({\bf A}_{0}) & \lapl \, & 0^{-}\cr
  \Lambda_5({\vec {\bf A}}\,) & \lapl \, & 1^{-}\cr
  \lambda_1({\widehat R})
  & \lapl \, & 0^{+}\cr \lambda_2({\widehat X}) & \lapl \,
  & 0^{-}\cr {\widetilde R} & \lapl \, & 2^{+}
\end{eqnarray}
In the case of a non-vanishing tensor torsion $^{(1)}T^{\alpha}\neq
0$, those terms deliver massive modes of spin state $2^{\pm }$ that
would combine with the corresponding spin $2^{+}$-mode of the
Riemannian curvature scalar ${\widetilde R}$. Because of the
complexity of these results, we will defer their presentation to
follow up work.

\subsection{Excitations of the gravitational field}\label{Sec.IVD}

We differentiate the Lagrangian (\ref{L_BHN'}) with respect to torsion
and curvature. Then, with the help of (\ref{excit}), we find the
translational excitation
\begin{eqnarray}\label{H_a}
H_{\alpha} & = & - \frac{1}{{\kappa}}\left( a_{2}\, ^{\star
(2)}T_{\alpha} + a_{3}\, ^{\star (3)}T_{\alpha}\right)\nonumber\\ && -
\frac{1}{{\kappa}}\,{{\sigma}_{2}}\left( ^{(2)}T_{\alpha} + \,
^{(3)}T_{\alpha}\right)
\end{eqnarray}
and the Lorentz excitation
\begin{eqnarray}\label{H_ab}
  H_{\alpha\beta} & = & \frac{1}{2{\kappa}}\left(
    a_{0}\,{\eta}_{\alpha\beta} + {b_{0}}\,{\vartheta}_{\alpha\beta}
  \right) \cr & & \cr & & +
  \frac{1}{{\varrho}}\left( w_{3}\, ^{\star (3)}R_{\alpha\beta} +
    w_{6}\, ^{\star (6)}R_{\alpha\beta}\right)\cr & & \cr & & +
  \frac{1}{2{\varrho}}{\mu}_{3}\left( ^{(3)}R_{\alpha\beta} +
    ^{(6)}R_{\alpha\beta}\right)\,,
\end{eqnarray}
respectively. Alternatively, the field excitations can be given in
terms of the field strengths in a more suitable and symmetric form
(${\cal V}={\cal V}_\a\vt^\a$ and ${\cal A}={\cal A}_\a\vt^\a$)
\begin{eqnarray}
  H_{\alpha} & = & \frac{1}{3\kappa}\left[ \left(
      a_{2}{\cal V}_{\mu}-{\sigma}_{2}{\cal
        A}_{\mu}\right){\eta}^{\mu}{}_{\alpha}
\right.\nonumber\\ && \hspace{7pt}\left.
 +\left(\sigma_{2}{\cal V}_{\mu}+a_{3}{\cal
        A}_{\mu}\right){\vartheta}^{\mu}{}_\a\right]\, , \label{Ha_I'} \\
  H_{\alpha\beta} & = & \left(
    \frac{a_{0}}{2\kappa}-\frac{w_{6}}{12\varrho}R-
    \frac{{\mu}_{3}}{24\varrho}X\right){\eta}_{\alpha\beta}\nonumber\\ &&
  \hspace{-8pt} + \left(\frac{b_{0}}{2\kappa}+\frac{w_{3}}{12\varrho}X
    -\frac{{\mu}_{3}}{24\varrho}R\right)
  {\vartheta}_{\alpha\beta}\label{H_ab'}\,.
\end{eqnarray}
{}From (\ref{Ha_I'}) we find in particular
\begin{equation}\label{trace_Ht}
  H_{\alpha}\wedge\, {\vartheta}^{\alpha} =
  \frac{1}{\kappa}\, ^{\star}\!\left(a_{2}{\cal V} - {\sigma}_{2}{\cal A}\right)\,.
\end{equation}


\subsection{Explicit form of field equations for the new
  Lagrangian}\label{Sec.IVE}

By substituting the excitations (\ref{H_a}), (\ref{H_ab}), and the
Lagrangian (\ref{L_BHN'}) into the gauge currents
(\ref{gaugecurrents}) and (\ref{gc2}), and then the latter two,
together with the excitations  (\ref{H_a}), (\ref{H_ab}), into the
field equations (\ref{first}) and (\ref{second}), we find the explicit
forms of the field equations: The first field equation reads,
\begin{widetext}\begin{eqnarray}\label{Sigma_1}
 && \left(\frac{a_{0}}{\kappa} -
  \frac{w_{6}}{6{\varrho}}R - \frac{{\mu}_{3}}{12{\varrho}}X\right)
G_{\alpha} + \frac{{\lambda}_{0}}{\kappa}\,{\eta}_{\alpha}
- \left( \frac{b_{0}}{\kappa} + \frac{w_{3}}{6{\varrho}}X -
  \frac{{\mu}_{3}}{12{\varrho}}R
\right)\, ^{\star}X_{\alpha} \cr & & \cr & &
+
\frac{1}{24\varrho}\left( w_{3}X^2 - w_{6}R^2 -
  {\mu}_{3}RX\right){\eta}_{\alpha}
+\frac{1}{3{\kappa}}D\left\{ ^{\star}\left[ \left(
      a_{2}{\cal V}-{\sigma}_{2}{\cal A}\right)\wedge\,
    {\vartheta}_{\alpha}\right]+\left( a_{3}{\cal
      A}+{\sigma}_{2}{\cal V}\right)\wedge\, {\vartheta}_{\alpha}\right\}\cr & &
\cr & &
+\frac{2a_{2}}{9\kappa}\left(
  {\cal V}_{\alpha}{\cal V}^{\b }-\frac{1}{4}{\cal V}^2
  {\delta}_{\alpha}^\b \right){\eta}_{\b }
+\frac{2a_{3}}{9\kappa}\left( {\cal A}_{\alpha}{\cal
    A}^{\b }-\frac{1}{4}{\cal
    A}^2{\delta}_{\alpha}^\b \right){\eta}_{\b }\cr & &
\cr & & +\frac{1}{\kappa} \left(
  e_{\alpha}\lrcorner\, ^{(1)}T^{\beta}\right)\wedge\left[ a_{2}\, ^{\star
    (2)}T_{\b} + a_{3}\, ^{\star (3)}T_{\b}{{ +}}\,
  {\sigma}_{2}\left( ^{(2)}T_{\b} + \,
    ^{(3)}T_{\b}\right)\right]
={\Sigma}_{\alpha}\,,
\end{eqnarray}
and the second field equation,
\begin{eqnarray}\label{Delta_anti}
  & & \frac{a_{0}}{2\kappa}\left(
    2\,^{\star (1)}T_{[\alpha}\wedge\, {\vartheta}_{\beta ]}
    -\frac{2}{3}{\cal V}\wedge\,
    {\eta}_{\alpha\beta} + \frac{1}{3}{\cal A}\wedge\,
    {\vartheta}_{\alpha\beta}\right) + \frac{b_{0}}{2\kappa}
  \left( 2\,^{(1)}T_{[\alpha}\wedge\,
    {\vartheta}_{\beta ]}-\frac{2}{3}{\cal V}\wedge\,
    {\vartheta}_{\alpha\beta} -
    \frac{1}{3}{\cal A}\wedge\, {\eta}_{\alpha\beta}\right)\cr & & \cr
  & & + \frac{1}{24{\varrho}}\left(
    2w_{3}dX-{\mu}_{3}dR\right)\wedge\,{\vartheta}_{\alpha\beta}
  -\frac{1}{24{\varrho}}\left(
    2w_{6}dR+{\mu}_{3}dX\right)\wedge\,{\eta}_{\alpha\beta} -\frac{1}{24{\varrho}}\left( 2w_{6}R+{\mu}_{3}X\right)T^{\g}\wedge\,
  {\eta}_{\alpha\beta\g}\cr & & \cr & & +
  \frac{1}{12{\varrho}}\left( 2w_{3}X-{\mu}_{3}R\right)T_{[\alpha}\wedge\, {\vartheta}_{\beta ]}
  -
  \frac{1}{3{\kappa}}\left[ a_{2}{\cal V}_{[\alpha}{\eta}_{\beta ]}
    -{\sigma}_{2}{\cal A}_{[ \alpha}{\eta}_{\beta ]} + \left(
      a_{3}{\cal A}+{\sigma}_{2}{\cal V}\right)\wedge\,
    {\vartheta}_{\alpha\beta}\right] = \tau_{\alpha\beta}\,.
\end{eqnarray}
\end{widetext}
The source of the second field equation is the material spin angular
momentum 3-form $\tau_{\alpha\beta}$. According to its
definition in (\ref{mattercurrents}), it is antisymmetric in $\a$ and
$\b$. It is related to the source of the first field equation, the
canonical energy-momentum 3-form of matter $\Sigma_\a$, via the
angular momentum law $D\tau_{\a\b}+\vt_{[\a}\wedge\Sigma_{\b]}=0$.


It may not be superfluous to look at the structures of the two field
equations and at the ways we ordered them. The first line of
(\ref{Sigma_1}) emerges {}from the curvature dependent pieces of the
gauge energy-momentum $E_\a$. We find, symbolically written, $\sim
\text{Einstein + cosmol.\ term + pseudoscalar curv + curv}^2$. If only
$a_0\ne 0$ and $b_0\ne 0$, we recover the left-hand-side of
(\ref{EC_pv1}); if only $a_0\ne 0$, we have just the EC-theory. The
second and third lines of (\ref{Sigma_1}) are of the following
structure: d torsion + torsion$^2$. {}From the point of view of gauge
theory, `d torsion' is the leading term, see $D H_\a$ in
(\ref{first}); the remaining `torsion$^2$' pieces collect the torsion
dependent parts of the gauge energy-momentum $E_\a$. Of course, also
the frame $e_\a$ and the coframe $\vt^\a$ feature in this equation
directly or indirectly via $\eta_\a=\,^\star\vt_\a$.

We ordered the second field equation (\ref{Delta_anti}) in a similar
way. In the first line we have the terms linear in torsion originating
{}from the gauge spin $E_{\a\b}$, see (\ref{gc2}); compare also as
special case the left-hand-side of (\ref{EC_pv2}). In the second and
third line, we have `d curv + curv $\times$ torsion'. The leading term
is `d curv', the rest arises {}from the differentiation process of `D
curv' with the help of (\ref{Deta}). Again the coframe $\vt^\a$ enters
explicitly or implicitly via $\vt_{\a\b}=\vt_\a\wedge\vt_\b$,
$\eta_{\a\b}=\,^\star(\vt_\a\wedge\vt_\b)$, and
$\eta_{\a\b\g}=\,^\star(\vt_\a\wedge\vt_\b\wedge\vt_\g)$.

Accordingly, the two field equations are now expressed in terms of the
torsion and the curvature of spacetime. In the sense of gauge theory
one may now want to insert the definitions of torsion, Eq.\
(\ref{torsion}), and curvature, Eq.\ (\ref{deccurv}) cum
(\ref{corv}). Then we would get second order quasilinear partial
differential equations (PDEs) in coframe and Lorentz-connection:
`$d\,^\star d\vt+\text{ lower order} \sim \text{energy-momentum}$' and
`$d\,^\star d\Gamma+\text{ lower order} \sim \text{spin}$'. However, our
experience on the search for exact solutions, in particular for
cosmological models, has shown that it is to be preferred to stay with
the well-behaved tensor-valued two-forms of torsion and curvature and
not to switch to the proper gauge variables coframe and
Lorentz-connection.

As we saw already, the Lagrangian (\ref{L_BHN'}) has a remarkable
symmetry which we will find also on the level of the excitations
(\ref{Ha_I'}) and (\ref{H_ab'}), the field equations (\ref{Sigma_1})
and (\ref{Delta_anti}), as well as on the level of the coupling
constants. Hence we can introduce the following tentative
correspondences (in four dimensions) between variables and parameters,
respectively,
\begin{equation}\label{symm_TA}
\begin{tabular}{|c|}\hline\cr
${\cal V} \lapl \, {\cal A}$ \cr\cr \hline
\end{tabular}\quad\quad
\begin{tabular}{|c|}\hline\cr
$a_{2} \lapl \, a_{3}$ \cr\cr \hline
\end{tabular}
\end{equation}
\begin{equation}\label{symm_wx}
\begin{tabular}{|c|}\hline\cr
$R \lapl \, X$ \cr $G_{\alpha}\lapl\, ^{\star}X_{\alpha}$\cr\cr
\hline
\end{tabular}\quad\quad
\begin{tabular}{|c|}\hline\cr
$w_{6} \lapl \, w_{3}$ \cr $a_{0}\lapl\, b_{0}$\cr\cr \hline
\end{tabular}
\end{equation}
\begin{equation}\label{symm_const}
\begin{tabular}{|c|}\hline\cr
${\sigma}_{2} \lapl \, {\mu}_{3}$ \cr\cr \hline
\end{tabular}
\end{equation}


\subsection{A consequence of a topological term}\label{Sec.IVF}

There is a subtlety present in the Lagrangian (\ref{L_BHN'}) and the
corresponding field equations (\ref{Sigma_1}) and
(\ref{Delta_anti}). Because of the Nieh-Yan identity
\cite{Nieh:1981ww}---see also \cite{PRs}, Eqs.(3.9.7) and
(B.4.15)---we have
\begin{equation}\label{NY}
d(\vt^\a\wedge T_\a)\equiv T^\a\wedge T_\a+R_{\a\b}\wedge\vt^{\a\b}
=T^\a\wedge T_\a-{}^\star\! X\,.
\end{equation}
The torsion square term can be expressed in its irreducible components
according to
\begin{eqnarray}\label{Tsquare}
  T^\a\wedge T_\a&=&\,^{(1)}T^\a\wedge{} ^{(1)}T_\a
  +2\,^{(2)}T^\a\wedge {} ^{(3)}T_\a\nonumber\\
  &=& \,^{(1)}T^\a\wedge{} ^{(1)}T_\a -\frac 23 {\cal A}\wedge{}^\star {\cal V}
  \,.
\end{eqnarray}
We substitute (\ref{Tsquare}) into the right-hand-side of the Nieh-Yan
identity (\ref{NY}) and find
\begin{eqnarray}\label{NY'}
  d(\vt^\a\wedge T_\a)=\,^{(1)}T^\a\wedge{} ^{(1)}T_\a -\frac 23
  {\cal A}\wedge{}^\star {\cal V}-{}^\star\! X\,.
\end{eqnarray}

For the sake of a neater argument let us first extend our parity mixed
PG Lagrangian (\ref{L_BHN'}) by including also the $\sigma_1$-term
{}from (\ref{parityodd}):
\begin{equation}\label{neuerL}
  \hat{V}=V_{\rm BHN}\,+\stackrel{\sigma_1}{V}
=V_{\rm BHN}\,{+\frac{\sigma_1}{\kappa}\, {}^{(1)}T^\alpha\wedge
  {}^{(1)}T_\alpha}\,.
\end{equation}
If we substitute (\ref{L_BHN'}) into (\ref{neuerL}), then we recognize
that we can recover {}from $\kappa\hat{V}$ the right-hand-side of the
Nieh-Yan identity (\ref{NY'}) for the specific coupling constants
\begin{equation}\label{choice}
  b_0=-2\,,\qquad\sigma_1=1\,,\qquad \sigma_2=2\,,
\end{equation}
all other constants, apart {}from $\kappa$ and $\varrho$, vanish. Since the
left-hand-side of (\ref{NY'}) is an exact form, the choice
(\ref{choice}) corresponds to a `null Lagrangian' with vanishing field
equations.

By the same token, we can add a multiple (say $\epsilon/\kappa$) of the
exact `topological' form $d(\vartheta^\alpha\wedge
T_\alpha)$ to $\hat{V}$. After some simple algebra, we find
\begin{align}\label{allerneuesterL}
  &\hspace{-12pt}\hat{V}+\frac{\epsilon}{\kappa}d(\vt^\a\wedge T_\a)=
  \frac{1}{\kappa}(\sigma_1+\epsilon)\, {}^{(1)}T^\alpha\wedge
  {}^{(1)}T_\alpha\nonumber\\
  &+ \frac{1}{2{\kappa}}\left[a_{0}{}^{\star}
  R+(b_{0}-2\epsilon){}^{\star}{X} -
  2{\lambda}_{0}{\eta}\right]\nonumber\\
  &+\frac{1}{6{\kappa}}\left[a_{2}{\cal V}\wedge{}^{\star}{\cal V} - a_{3}{{\cal
        A}\wedge{} ^{\star}{\cal A}} -2({\sigma}_{2}+2\epsilon){ {\cal A}\wedge{}
      ^{\star}{\cal V}}\right]\nonumber\\
  &-\frac{1}{24{\varrho}}\left( w_{6}R\,^{\star}R- w_{3} {X\,^{\star} X}
    +
    {\mu}_{3}R\,^\star X\right)\,.
\end{align}

This is equivalent to certain changes in the parameters of our action
(\ref{neuerL}), specifically
\begin{equation} \sigma_1\to \sigma_1+\epsilon,\quad b_0\to
  b_0-2\epsilon, \quad \sigma_2\to
  \sigma_2+2\epsilon.\label{alphatrans}
\end{equation}
{}From this we can infer that the field equations cannot depend
on the parameters  $\sigma_1$, $b_0$, $\sigma_2$ by themselves, but
rather must depend on these parameters only through certain
combinations which are {\it invariant} under this transformation, such as
\begin{equation}\label{comb}b_0+2\sigma_1\,,\qquad
  2\sigma_1-\sigma_2\,, \qquad b_0+\sigma_2\,.
\end{equation}

The two field equations of the Lagrangian $\hat{V}$ are found via
$\stackrel{\sigma_1}{H}_\a=-2\sigma_1\,^{(1)}T_\a/\kappa$ as
\begin{align}\label{first'}
  & \text{l.h.s.\ of }(\ref{Sigma_1})-\frac{2\sigma_1}{\kappa}\left[
    D\,^{(1)}T_\a\right.\nonumber\\ &\hspace{20pt}\left.+ (e_\a\lrcorner
    \,^{(1)}T^\b)\wedge(\,^{(2)}T^\b+\,^{(3)}T_\b )\right]
  ={\Sigma}_{\alpha}\,,
\end{align}
\begin{align}
  \text{l.h.s. of
  }(\ref{Delta_anti})-\frac{2\sigma_1}{\kappa}\,\vt_{[\a}\wedge\,^{(1)}T_{\b]}
  = \tau_{\alpha\beta}\,.\label{second'}
\end{align}
Specifically, using the first Bianchi identity,
\begin{equation}^{\star\!} X_\alpha=
  R_{\beta\alpha}\wedge\vartheta^\beta\equiv DT_\alpha=
  D({}^{(1)}T_\alpha+{}^{(2)}T_\alpha+{}^{(3)}T_\alpha)\,,
\end{equation}
and the expressions (\ref{trator}) and (\ref{axitor}) for the
irreducible pieces, one can find that the $b_0$, $\sigma_1$, and
$\sigma_2$ terms on the left-hand-side of the first field equation
(\ref{first'}) add up to
\begin{eqnarray} &&\hspace{-8pt}-\frac{1}{\kappa} D\left[(b_0+2\sigma_1)
    {}^{(1)}T_\alpha\nonumber
    +(b_0+\sigma_2)\left({}^{(2)}T_\alpha+{}^{(3)}T_\alpha\right)\right]\\
  &&+\frac{1}{\kappa}(2\sigma_1-\sigma_2)
  (e_\alpha\lrcorner{}^{(1)}T^\beta)\wedge\left({}^{(2)}T_\beta
    +{}^{(3)}T_\beta\right)\,;\label{Jim1st}
\end{eqnarray}
similarly, for the left-hand-side of the second field equation
(\ref{second'}) we have
\begin{equation}
  \frac1{\kappa}(b_0+2\sigma_1){}^{(1)}T_{[\alpha}\wedge
    \vartheta_{\beta]}
    -\frac{1}{6\kappa}(b_0+\sigma_2)(2 {\cal V}\wedge\vartheta_{\alpha\beta}
    +{\cal A}\wedge\eta_{\alpha\beta}).\label{Jim2nd}
\end{equation}

There are several points worth noting: (i) As expected, the parameters
occur only in the invariant combinations (\ref{comb}). (ii) All the
$\sigma_1$ terms are proportional to $^{(1)}T^\mu$. (iii) For
solutions such that $^{(1)}T^\mu$ vanishes, the equations contain
the parameters only in the combination $b_0+\sigma_2$.

Having obtained these insights, it is no longer necessary to keep the
rather complicated $\sigma_1$ term.  Exploiting the freedom to choose
a suitable $\epsilon$ in (\ref{alphatrans}), namely $\epsilon=-\sigma_1$, we
can, without loss of generality, drop the $\sigma_1$-term altogether
and return to our model Lagrangian (\ref{L_BHN'}) with the two field
equations (\ref{Sigma_1}) and (\ref{Delta_anti}).


\section{Friedman cosmologies with propagating modes of the
  Lorentz-connection}\label{Sec.V}

Since the early 1970s cosmological models for EC and PG have been
developed, see Kopczy\'nski \cite{Wojtek73}, Trautman
\cite{Andrzej73}, Tafel \cite{Tafel}, Kuchowicz
\cite{Kuchowicz1,Kuchowicz2}, Kerlick \cite{Kerlick:1976}, and others
\cite{Nara,Canale}, to name a few. Minkevich et al.\ \cite{Minkevich}
developed the subject in a series of papers. A report on the status of
the subject was given by Puetzfeld \cite{Puetzfeld:2004yg}.

For our new Lagrangian we follow these procedures and search for FLRW
type of cosmological models.

\subsection{Homogeneous and isotropic coframe and torsion}\label{Sec.VA}

Assuming a homogeneous and isotropic scenario, the {\it orthonormal coframe}
for a Friedman cosmos is
\begin{eqnarray}\left.\begin{array}{lll}
{\vartheta}^{0} & = & dt\, \vspace{5pt} \\ {\vartheta}^{1} & = &\displaystyle
\frac{a(t)dr}{\sqrt{1-kr^{2}}}\,   \vspace{5pt}\\  {\vartheta}^{2} & =
& a(t)rd{\theta}\,   \vspace{5pt}\\  {\vartheta}^{3} & = &
a(t)r{\sin\theta}d{\phi}\, \end{array}\right\}\,,\label{cofr}
\end{eqnarray}
with the metric
\begin{eqnarray}
\hspace{-20pt}
g&\hspace{-5pt}= \hspace{-5pt}&
  -{\vartheta}^{0}\otimes {\vartheta}^{0} +
  \sum\limits_{a=1}^{3} {\vartheta}^{a}\otimes {\vartheta}^{a}\cr &
  \hspace{-5pt} = \hspace{-5pt}&
 -dt^2\! +\! \frac{a^2(t)}{1-kr^2}
  dr^2\! +\! a^2(t)r^2\!\left(
    d{\theta^2} + {\sin^2{\theta}}d{\phi}^2\right)\,,\cr&&
\end{eqnarray}
where $a(t)$ is the expansion factor and $k$ the curvature index.

Now we can compute, up to antisymmetry, the nonvanishing components of the
Riemannian connection ($a,b,c,...=1,2,3$ are spatial anholonomic
(frame) indices):
\begin{eqnarray}
  \left.\begin{array}{llr}
      \widetilde{\Gamma}_{a}{}^{0} & = &\displaystyle
      \frac{a'(t)}{a(t)}\vt^a
      \cr & & \cr
      \widetilde{\Gamma}_{2}{}^{1} & = & -\displaystyle
      \frac{\sqrt{1-kr^2}}{a(t)r}\,
      {\vartheta}^{2}\,  \cr & & \cr
      \widetilde{\Gamma}_{3}{}^{1}& = & -\displaystyle
      \frac{\sqrt{1-kr^2}}{a(t)r}\, {\vartheta}^{3}\,  \cr & & \cr
      \widetilde{\Gamma}_{3}{}^{2}& = & - \displaystyle
      \frac{{\cot {\theta}}}{a(t)r}\,
      {\vartheta}^{3}\, \end{array}\right\}\,.\label{Riemconn}
\end{eqnarray}
Like in Einstein's theory, the temporal rate of change of the
expansion factor $a(t)$ determines the Hubble function
\begin{equation}\label{Hubble}
  H(t):=a'(t)/a(t)\,.
\end{equation}
By differentiation and elimination of $a'(t)$, we find
\begin{equation}\label{Hubble'}
H'(t)+H^2(t)=a''(t)/a(t)\,.
\end{equation}
This determines the Riemannian sector of spacetime.

The most general torsion compatible with homogeneity and isotropy
can be characterized by two independent functions $u(t)$ and
$v(t)$, see Goenner \& M\"uller-Hoissen \cite{Goenner} and Baekler
\cite{BaeklerDr}. We will choose for the torsion the
parametrization
\begin{eqnarray}\label{tor}
  \left.\begin{array}{lll}T^{0} & = & 0\,
      \cr T^{1} & = & u(t){\vartheta}^{01}
      +v(t){\vartheta}^{23} \cr
      T^{2} & = & u(t){\vartheta}^{02}
      +v(t){\vartheta}^{31} \cr
      T^{3} & = & u(t){\vartheta}^{03}
      +v(t){\vartheta}^{12}
    \end{array}\right\}\,.
\end{eqnarray}
The irreducible decomposition of $T^{\alpha}$ implies a vanishing
tensor piece, whereas the vector and axial vector pieces survive:
\begin{equation}
 ^{(1)}T^{\alpha}  =  0\, ,\> ^{(2)}T^{\alpha}  = u\left(
   \begin{matrix}
     0\cr {\vartheta}^{01}\cr {\vartheta}^{02}\cr {\vartheta}^{03}
   \end{matrix}\right)\, ,\>
 ^{(3)}T^{\alpha}  = v\left(
\begin{matrix}
  0\cr {\vartheta}^{23}\cr {\vartheta}^{31}\cr {\vartheta}^{12}
\end{matrix}\right)\,.
\end{equation}
This yields for the corresponding 1-forms in (\ref{trator}) and
(\ref{axitor})
\begin{equation}\label{T_A}
  {\cal V} = -3u(t){\vartheta}^{0}\, , \quad {\cal A} =
  -3v(t){\vt}^{0}\, .
\end{equation}
Incidentally, the purely spatial part of the torsion
$^{(3)}T^{\alpha}$ corresponds to Cartan's spiral staircase
\cite{Cartan1922,LazH2010} with a time dependent pitch of the
spiral. As such, one can easily visualize it, see \cite{LazH2010}.

By simple algebra, we can calculate the contortion $K_{\alpha\beta}$
in terms of the torsion, see \cite{PRs},
\begin{equation}\label{contortion}
 K_{\alpha\beta}= e_{[\alpha}\lrcorner T_{\beta ]} -{1\over 2}\,
( e_{\alpha}\lrcorner e_{\beta}\lrcorner T_{\gamma})\vartheta^{\gamma}\,.
\end{equation}
We find then the Riemann-Cartan connection according to
$\Gamma_{\a\b}=\widetilde{\Gamma}_{\a\b}-K_{\a\b}$ or, explicitly,
\begin{eqnarray}\label{RC_con}\left.\begin{array}{llr}
{\Gamma}_{a}{}^{0} & = &\displaystyle
 \left[H(t)-u(t)\right]{\vartheta}^{a}\,\cr
{\Gamma}_{a}{}^{b} & = &\displaystyle \widetilde{\Gamma}_a{}^b+
\frac{1}{2}v(t){\epsilon}_{ca}{}^{b}{\vartheta}^{c}\,\end{array}\right\}\,.
\end{eqnarray}

\subsection{Irreducible pieces of the curvature}\label{Sec.VB}

Having now coframe, connection, and torsion at our disposal, we
can immediately calculate the different pieces of the curvature
2-form. We introduce the Hubble function $H(t)$ (\ref{Hubble}) and
find straightforwardly
\begin{eqnarray}
  ^{(1)}R_{\alpha\beta} & = & 0\, , \\ 
  ^{(2)}R_{\alpha\beta}\label{WW2}
  & = &\hspace{-5pt}
  \frac{1}{4}\!\!\left\{v(t)[H(t)-2u(t)]\right.\nonumber\\ &&
\left.\hspace{10pt}-v'(t)\right\}\hspace{-3pt}\left(
    \begin{matrix}0 & {\vartheta}^{23} & -{\vartheta}^{13} &
      {\vartheta}^{12}\cr \diamond & 0 &
      -{\vartheta}^{03} &  {\vartheta}^{02}
      \cr \diamond & \diamond & 0 & -{\vartheta}^{01}\cr
      \diamond & \diamond & \diamond &
      0\end{matrix}\right),\\
  \cr
 ^{(3)}R_{\a\b}&=&-\frac{1}{12}\,X(t)\eta_{\a\b} 
 \, , \\ \label{WW4}
 ^{(4)}R_{\alpha\beta} & = &\frac{1}{2}
 \left[H'(t)-u'(t)+H(t)u(t)-u^2(t)\right.\nonumber \\ &&\left.
\hspace{-18pt} +\frac{1}{4}v^2(t)-\frac{k}{a^2(t)}\right]
\left(
\begin{matrix}
  0 & {\vartheta}^{01} & {\vartheta}^{02} & {\vartheta}^{03}\cr
  \diamond & 0 & {\vartheta}^{12} & {\vartheta}^{13}\cr \diamond &
  \diamond & 0 & {\vartheta}^{23}\cr \diamond & \diamond & \diamond &
  0
\end{matrix}\right), \\ 
^{(5)}R_{\alpha\beta} & = & 0\, , \\
^{(6)}R_{\a\b}&=&-\frac{1}{12}\,R(t)\vt_{\a\b}\,, 
\end{eqnarray}
with the (pseudo)-scalar functions
\begin{eqnarray}\label{XX}
  X(t)&=&-3\left\{v'(t)+v(t)[3H(t)-2u(t)]\right\}\,,\cr && \\  \label{WW}
  R(t)&=&6\left\{[H'(t)-u'(t)]+H(t)[H(t)-u(t)]\right.\nonumber\\
  &&\left.+[H(t)-u(t)]^2
    -\frac{1}{4}v^2(t)+\frac{k}{a^2(t)}\right\}\,.
\end{eqnarray}
The matrices in (\ref{WW2}) and (\ref{WW4}) are antisymmetric,
respectively. Therefore, we indicated those matrix elements with a
diamond symbol that are, because of this antisymmetry, redundant.

Since we chose as our variables curvature and torsion, the relations
between curvature and torsion provided by the nonvanishing irreducible
pieces and, in particular, by (\ref{XX}) and (\ref{WW}), are vital
information in our search for exact solutions.

\subsection{A spinless perfect fluid as model of matter}\label{Sec.VC}

The sources of the two field equations are the energy-momentum current
$\Sigma_\a$ and the spin current $\tau_{\a\b}$. These three-forms we
represent as tensor-valued zero-forms according to
\begin{equation}\label{Sigma_a}
{\Sigma}_{\alpha}={\cal T}_{\alpha}{}^{\mu}{\eta}_{\mu}\,,
\qquad \tau_{\a\b}= S_{\a\b}{}^\mu\eta_\mu\,;
\end{equation}
the reciprocal relations read
\begin{eqnarray} {\cal T}_{\alpha}{}^{\beta}=\,^{\star}\left(
    {\Sigma}_{\alpha}\wedge {\vartheta}^{\beta}\right) \,,\qquad
 {S_{\a\b}{}^\g}=\,^{\star}\left(
    {\tau}_{\alpha\b}\wedge {\vartheta}^{\g}\right)\,.
\end{eqnarray}

In the following we will only consider matter models with vanishing
spin, ${\tau}_{\alpha\beta}=0$. This simplifying assumption, which may
be justified for spherical symmetry, certainly has to be dropped in a
more advanced stage of our model building.

Due to the Friedman (or FLRW) symmetry of our cosmological model, the
energy-momentum tensor must have the spinless perfect fluid form
\begin{equation}\label{p_fluid_1}
{\cal T}_{\alpha}{}^{\beta} = [{\rho}(t)+p(t)]U_{\alpha}U^{\beta}
+ p(t){\delta}_{\alpha}^{\beta}\,,
\end{equation}
where $\rho=\rho(t)$ is the energy density, $p=p(t)$ the pressure, and
$U_\a$ the four-velocity of the fluid, with the normalization
$U^{\alpha}U_{\alpha}=-1$. Because of the symmetry requirements, we
only have the dependencies ${\rho}(t)$ and ${p(t)}$. In a comoving
reference system, we have $U^{\alpha}={\delta}^{\alpha}_{0}\, ,
U^{0}=1$, and
\begin{equation}\label{p_fluid_2} {\cal T}_{\alpha}{}^{\beta}
  \stackrel{*}{=} {\rm diag}\,\left( -{\rho}(t)\, , p(t)\, ,
    p(t)\, , p(t)\right)\,.
\end{equation}
{}From (\ref{p_fluid_1}) we deduce for the trace
\begin{equation}\label{trace_cal_T}
  {\cal T}_{\alpha}{}^{\alpha}=-{\rho}(t)+3p(t)
\end{equation}
\def\calT_d{\nearrow\hspace{-13pt}
{\cal T}}
and for the trace-free part
\begin{eqnarray}
 {\cal T}^\dagger{}\hspace{-4pt}_{\alpha}{}^{\beta}&=& [\rho(t)+p(t)]\left(U_\a
    U^\b+\frac 14 \d_\a^\b\right) \nonumber\\ &\stackrel{*}{=}&
    \frac{1}{4}\left[{\rho}(t)+p(t)\right]\,{\rm diag}\, ( -3,1,1, 1)\,.
\end{eqnarray}

\subsection{Differential equations for torsion and curvature}\label{Sec.VD}

According to the FLRW-symmetry requirements, the first field
equation (\ref{Sigma_1}) as a vector-valued 3-form, has only two
(algebraically) independent components and in the same manner the
second field equation (\ref{Delta_anti}) as an antisymmetric
tensor-valued 3-form also has only two independent components. 

\subsubsection{First field equation}\label{Sec.VD1}

We substitute into the first field equation (\ref{Sigma_1}) the
coframe (\ref{cofr}), the torsion (\ref{tor}), and the expansion
factor via (\ref{Hubble}) and (\ref{Hubble'}), but leave $R(t)$ and
$X(t)$ as they are. Then we find as independent non-vanishing
equations only the components $\rho(t)=e_1\lrcorner [e_2\lrcorner (
e_3\lrcorner \Sigma_0)]$ and $p(t)=e_0\lrcorner[ e_2\lrcorner
(e_3\lrcorner\Sigma_1)]$:
\begin{widetext}
\begin{eqnarray}\label{f_1}
{\kappa}{\rho}(t) & = &
\frac{1}{2}\left[a_{0}R(t)+b_{0}X(t)\right]+3a_{0}\left\{[u(t)
  -H(t)]^{\prime}+H(t)[u(t)-H(t)]\right\}
+\frac{3}{2}b_{0}\left[v^{\prime}(t)+H(t)v(t)\right]\cr & & \cr &
&
-\frac{3}{2}\left\{a_{2}u(t)\left[u(t)-2H(t)\right]-a_{3}v^2(t)\right\}
+3{\sigma}_{2}v(t)\left[u(t)-H(t)\right]\cr
& & \cr & &
-\frac{\kappa}{4{\varrho}}[2w_{6}R(t)+{\mu}_{3}X(t)]\left\{[u(t)-H(t)]^{\prime}+
  H(t)[u(t)-H(t)]\right\}\cr & & \cr & &
-\frac{\kappa}{8{\varrho}}[{\mu}_{3}R(t)-2w_{3}X(t)]\left[v^{\prime}(t)
  +H(t)v(t)\right]\cr
& & \cr & &
-\frac{\kappa}{24{\varrho}}\left[w_{6}R^2(t)+{\mu}_{3}R(t)X(t)
  -w_{3}X^2(t)\right]-{\lambda}_{0}
\end{eqnarray}
and
\begin{eqnarray}\label{f_2}
{\kappa}p(t) & =
&-\frac{1}{2}\left[a_{0}R(t)+b_{0}X(t)\right]+a_{0}\left[H^{\prime}(t)+3H^2(t)
  +\frac{2k}{a^2(t)}\right]-(a_{0}+a_{2})u^{\prime}(t)
+\frac{1}{2}(2{\sigma}_{2}-b_{0})v^{\prime}(t)\cr & & \cr & & -
(5a_{0}+2a_{2})H(t)u(t) -
\frac{1}{2}(5b_{0}-4{\sigma}_{2})H(t)v(t) +
\frac{1}{2}(4a_{0}+a_{2})u^2(t) + (2b_{0}-{\sigma}_{2})u(t)v(t)\cr
& & \cr & & - \frac{1}{2}(a_{0}+a_{3})v^2(t) + {\lambda}_{0}\cr &
& \cr & & -
\frac{\kappa}{12{\varrho}}[2w_{6}R(t)+{\mu}_{3}X(t)]\left[H^{\prime}(t)
  +3H^2(t)+\frac{2k}{a^2(t)}-u^{\prime}(t)-5H(t)u(t)+2u^2(t)
  -\frac{1}{2}v^2(t)\right]
\cr & & \cr & & +
\frac{\kappa}{24{\varrho}}[{\mu}_{3}R(t)-2w_{3}X(t)]\left[
  v^{\prime}(t)+5H(t)v(t)-4u(t)v(t)\right]\cr & & \cr & &  
+\frac{\kappa}{24{\varrho}}\left[w_{6}R^2(t)+{\mu}_{3}R(t)X(t)
  -w_{3}X^2(t)\right]\,.
\end{eqnarray}
A further relation between the fluid density $\rho(t)$ and the
pressure $p(t)$ can be gained by taking the trace
$\vartheta^\a\wedge\Sigma_\a$ of the first field equation or,
equivalently, by computing $\rho(t)-3 p(t)$ {}from (\ref{f_1}) and
(\ref{f_2}).  However, in order to find a compact expression, we
resolve (\ref{XX}) with respect to $v'(t)$ and (\ref{WW}) with respect
to $H'(t)$. If we substitute these expressions, we find\footnote{That
  on the right-hand-side only a derivative with respect to the torsion
  can appear can be seen from the general structure of this
  expression: $ \vt^\a\wedge\Sigma_\a=(\rho-3p)\eta=-d(\vt^\a\wedge
  H_\a) -4V-T^\a\wedge H_\a -2R_\a{}^\b\wedge H^\a{}_\b\,.$ }
\begin{eqnarray}\label{firstcombined}
{\kappa}\left[ {\rho}(t)-3p(t)\right] & = & a_{0}R(t) +
(b_{0}+{\sigma}_{2})X(t) +
3a_{2}\left[u^{\prime}(t)-u^2(t)+3H(t)u(t)\right]+3a_{3}v^2(t)-4{\lambda}_{0}
\cr &\hspace{-7pt}=\hspace{-7pt}&\frac{1}{2}(2a_{0}-a_{2})R(t) + (b_{0}
+{\sigma}_{2})X(t)\cr&&+\frac{3}{4}(4a_{3} -a_{2})v^2(t)
+3a_{2}\left( H^{\prime}(t)+2H^2(t)+\frac{k}{a^2(t)}\right) -4{\lambda}_{0}\,.
\end{eqnarray}

Let us now get back to the first field equation (\ref{f_1}) with
(\ref{f_2}). One strategy is to eliminate the time derivative $H'(t)$
of the Hubble function by a suitable linear combination of (\ref{f_1})
and (\ref{f_2}). Accordingly, we put the linear combination $
\left\{e_1\lrcorner [e_2\lrcorner (e_3\lrcorner \Sigma_0)]+3e_0
  \lrcorner
  [e_2\lrcorner( e_3\lrcorner\Sigma_1)]-[\rho(t)+3p(t)]\right\} $ to
zero and, isolating the derivatives of the torsion functions, we find
\begin{eqnarray}\label{up-vp}
  -3\left[ a_{2}u^{\prime}(t)-{\sigma}_{2}v^{\prime}(t)\right] & = &
  {\kappa}\left[{\rho}(t)+3p(t)\right] 
  +3H(t)\left[ a_{2}u(t)-{\sigma}_{2}v(t)\right] 
  +  a_{0}R(t)+b_{0}X(t)\cr&&\cr&& +
  6b_{0}v(t)\left[H(t)-u(t)\right] +
  \frac{\kappa}{2{\varrho}}\left[2w_{3}X(t)-{\mu}_{3}R(t)\right]v(t)
  \left[H(t)-u(t)\right]
  \cr & & \cr & &  -6a_{0}\left\{ [H(t)-u(t)]^2 
    - \frac{1}{4}v^2(t)+\frac{k}{a^2(t)}
  \right\}\cr & & \cr
  & & + \frac{\kappa}{2{\varrho}}\left[{\mu}_{3}X(t)+2w_{6}R(t)\right]
  \left\{\left[H(t)-u(t)\right]^2
    -\frac{1}{4}v(t)^2+\frac{k}{a^2(t)}\right\}\cr
  &&\cr&&-2{\lambda}_{0}-\frac{\kappa}{12\varrho}
  \left[w_6R^2(t)+\mu_3R(t)X(t)-w_3X^2(t) \right]\,.
\end{eqnarray}
Alternatively, the last relation (\ref{up-vp}) can also be
expressed as
\begin{eqnarray}\label{alter_uv}
\hspace{-30pt}  -\frac{3}{a(t)}\!\!& &\!\! \frac{d}{dt}\left\{a(t)
\left[a_{2}u(t)-{\sigma}_{2}v(t)\right]\right\}  = 
{\kappa}\left[{\rho}(t)+3p(t)\right] +  a_{0}R(t)+b_{0}X(t) +
6b_{0}v(t)\left[H(t)-u(t)\right]\cr & & \cr & & +
\frac{\kappa}{2{\varrho}}\left(2w_{3}X(t)-{\mu}_{3}R(t)\right)v(t)
\left[H(t)-u(t)\right]
-6a_{0}\left\{ [H(t)-u(t)]^2 -
\frac{1}{4}v^2(t)+\frac{k}{a^2(t)}\right\}
\cr & & \cr & & +
\frac{\kappa}{2{\varrho}}\left[{\mu}_{3}X(t)+2w_{6}R(t)\right]
\left\{\left[H(t)-u(t)\right]^2
-\frac{1}{4}v(t)^2+\frac{k}{a^2(t)}\right\}-2{\lambda}_{0}\cr & & \cr & &
-\frac{\kappa}{12\varrho}
    \left[w_6R^2(t)+\mu_3R(t)X(t)-w_3X^2(t) \right]\,.
\end{eqnarray}
\end{widetext}
This equation suggests to impose the interrelationship
\begin{equation}\label{constraint'}
  u(t)=\sigma\,
  v(t)\,,\qquad\text{with}\qquad \sigma:=\frac{\sigma_2}{a_2}\,,
\end{equation}
between the two torsion pieces as a simple special case; a further
possible choice could be
\begin{equation}\label{constraint2}
u(t)-\sigma v(t)=\frac{c}{a(t)}\,, \quad c={\rm constant}\neq 0\,.
\end{equation}
Then the left-hand-sides of (\ref{up-vp}) and (\ref{alter_uv}) vanish
and we find a purely algebraic equation in the variables
$\{v(t)\text{(or $u(t)$)}, R(t),X(t),H(t),\rho(t),p(t)\}$. We will
defer the study of these two alternatives to future work.

We can also manipulate the first field equation (\ref{f_1}) with
(\ref{f_2}) in a different way in order to arrive at algebraic
relations. We can resolve (\ref{WW}) and (\ref{XX}) with respect
to $u'(t)$ and $v'(t)$ and substitute these expressions into
(\ref{f_1}) and (\ref{f_2}), respectively.

After eliminating the derivatives of the torsion we arrive at
\begin{widetext}
\begin{eqnarray}\label{density}
{\kappa}{\rho}(t) & = & -{3}m^+u(t)\left[
  2H(t)-u(t)\right] - \frac{3}{4}m^-v^2(t) +
3a_{0}\left(H^2(t)+\frac{k}{a^2(t)}\right)\cr&&\cr&& -
3m^\times\left[H(t)-u(t)\right]v(t)
-\frac{{\kappa}}{4{\varrho}}\left[{\mu}_{3}X(t)+2w_{6}R(t)\right]\left\{
  [H(t)-u(t)]^2 - \frac{1}{4}v^2(t)+\frac{k}{a^2(t)}\right\}\cr&&\cr&& +
\frac{{\kappa}}{4{\varrho}}\left[{\mu}_{3}R(t)-2w_{3}X(t)\right]
v(t)[H(t)-u(t)]\cr
& & \cr 
& & 
+ \frac{{\kappa}}{24{\varrho}}\left[
  w_{6}R^2(t)-w_{3}X^2(t)+{\mu}_{3}R(t)X(t)\right] - {\lambda}_{0}\,
\end{eqnarray}
and
\begin{eqnarray}\label{pressure}
{\kappa}p(t) & = & -\frac{1}{3}m^+R(t) -
\frac{1}{3}m^\times X(t) +2(m^+-a_{0})H^{\prime}(t) +
m^\times v(t)\left[u(t)-H(t)\right]\cr
& & \cr & &+\frac{1}{4}m^- v^2(t) +
a_{0}\left\{\left[H(t)-u(t)\right]^2-\frac{1}{2}v^2(t)
  +\frac{k}{a^2(t)}\right\}\cr &&\cr && 
+2(m^+-a_{0})\left\{H(t)\left[2H(t)-u(t)\right]
  +\frac{1}{2}u^2(t)-\frac{1}{4}v^2(t)+\frac{k}{a^2(t)}\right\}
\cr & & \cr & &
+\frac{\kappa}{12{\varrho}}\left[{\mu}_{3}R(t)-2w_{3}X(t)\right]v(t)\left[H(t)-u(t)\right]\cr
& & \cr & &
-\frac{\kappa}{12{\varrho}}\left[{\mu}_{3}X(t)+2w_{6}R(t)\right]\left\{
  \left[H(t)-u(t)\right]^{2}-\frac{1}{4}v^2(t)+\frac{k}{a^2(t)}\right\}\cr
& & \cr & & + \frac{\kappa}{72{\varrho}}\left[
  w_{6}R^2(t)-w_{3}X^2(t)+{\mu}_{3}R(t)X(t)\right] + {\lambda}_{0}
\,.
\end{eqnarray}
Inspecting the equations there are the following dependencies,
\begin{eqnarray}
{\rho}(t) & = &
{\rho}\left[a(t),H(t),u(t),v(t),R(t),X(t)\right]\,,\cr p(t) & = &
p\left[a(t),H(t),H^{\prime}(t),u(t),v(t),R(t),X(t)\right]\,,
\end{eqnarray}
that is, the density depends only algebraically on the variables
whereas the pressure, besides non linear terms, contains only the
derivative of $H(t)$. Thus, also this general case belongs to the
class of {\em descriptor systems}, that is, to the
differential-algebraic systems.

\subsubsection{Second field equation}\label{Sec.VD2}

Similarly, also for the second field equation (\ref{Delta_anti}),
we find only two independent components. Both vanish by the
assumption of vanishing matter spin ${\tau}_{\alpha\beta}$. Thus,
\begin{align}
  \kappa\left\{2 w_{6}R^{\prime}(t)\right.&\left.
    +\mu_{3}X^{\prime}(t)+ \left[{\mu}_{3}v(t) +4w_{6}u(t)\right]R(t)
    + 2 \left[{\mu}_{3}u(t)-w_{3}v(t)\right]X(t)\right\}\cr &
  -12\varrho\left[2m^+ u(t)+m^\times v(t)\right]
   = 24\kappa \varrho\, e_0\lrcorner
  [e_2\lrcorner(e_3\lrcorner\tau_{01})]& = 0\,,
  \label{s_1x}\\
  & \cr
  \kappa\left\{{\mu}_{3}R^{\prime}(t)\right.&\left.-2w_{3}X^{\prime}(t)
    +2[{\mu}_{3}u(t)-w_{6}v(t)]R(t)
    -[4w_{3}u(t)+{\mu}_{3}v(t)]X(t)\right\}\cr &
  -12\varrho\left[ 2m^\times u(t)
    -m^- v(t)\right] = 24 \kappa \varrho\,
  e_0\lrcorner[e_2\lrcorner(e_3\lrcorner\tau_{23})]& =0 \, .\label{s_2x}
\end{align}
These are two ordinary linear differential equations (ODEs) of
first order in the curvature components $R(t)$ and $X(t)$. By
suitable linear combinations, we can uncouple the first
derivatives of these equations. We add $2w_3\times$ Eq.\
(\ref{s_1x}) to $\mu_3\times$ Eq.\ (\ref{s_2x}) and find, provided
$\Delta=\mu_3^2 +4w_3w_6\ne 0$, see (\ref{BBB}),
\begin{align}\label{WStrich}
  R'(t)&+2\left(u(t)+\mu_3\frac{w_3-w_6}{\Delta}v(t)\right)
  R(t) -\frac{\mu_3^2+4w_3^2}{\Delta}v(t)X(t)\cr
  &-\frac{12\varrho/\kappa}{\Delta}\left[2\left(2m^+ w_3
        +m^\times \mu_3\right)u(t)
    +\left(-m^- \mu_3+2m^\times w_3 \right)v(t)\right] \quad =0\,.
\end{align}
Similarly, we add $\mu_3\times$ Eq.\ (\ref{s_1x}) to $-2w_6\times$ Eq.\
(\ref{s_2x}) and find, provided $\mu_3^2 +4w_3w_6\ne 0$,
\begin{align}\label{XStrich}
  X'(t)&+2\left(u(t)-\mu_3\frac{w_3-w_6}{\Delta}v(t)\right)
  X(t) +\frac{\mu_3^2+4w_6^2}{\Delta}v(t)R(t)\cr
  &-\frac{12\varrho/\kappa}{\Delta}\left[\left(2m^+ \mu_3
        -4m^\times w_6\right)u(t)
    +\left(2m^- w_6+m^\times \mu_3 \right)v(t)\right]\quad =0\,.
\end{align}

Let us introduce, by also using (\ref{emms}), the following
abbreviations
\begin{eqnarray}\label{omega_01}
{\omega}_{0}^2  :=  \frac{4\varrho}{\kappa}\,\frac{
  2m^-w_{6}+m^\times\mu_{3}}{\Delta}\,,\qquad
{\omega}_{1}^2  := 
\frac{4\varrho}{\kappa}\,\frac{ 
  2m^+w_{3}+m^\times{\mu}_{3}}{\Delta}\,,\end{eqnarray}
\begin{eqnarray}\label{omega_23}
{\omega}_{2}^2  := 
 \frac{4\varrho}{\kappa}\,\frac{
{m^-\mu}_{3}-2m^\times w_{3}}{\Delta}\,,\qquad 
{\omega}_{3}^2  :=  
 \frac{4\varrho}{\kappa}\,\frac{
2m^+{\mu}_{3}-4m^\times w_{6}}{\Delta}\,.
\end{eqnarray}
The constants $\omega_0^2$ and $\omega_1^2$ are of even parity,
whereas $\omega_2^2$ and $\omega_3^2$ are of odd parity. This allows
us to give the components of the second field equation (\ref{WStrich})
and (\ref{XStrich}) a more compact and transparent form,
\begin{eqnarray}\label{sec_nonlinear}
\hspace{-14pt}R^{\prime}(t) & =& 6{\omega}_{1}^2u(t)-3{\omega}_{2}^2v(t) -
2u(t)R(t)
+\frac{v(t)}{\Delta}\left[\left(
{\mu}_{3}^2+4w_{3}^2\right)X(t) -
2{\mu}_{3}(w_{3}-w_{6})R(t)\right],\\ & & \cr \hspace{-14pt}X^{\prime}(t) & = &
 3{\omega}_{3}^2u(t)+3{\omega}_{0}^2v(t) - 2u(t)X(t)
- \frac{v(t)}{\Delta}\left[\left(
{\mu}_{3}^2+4w_{6}^2\right)R(t) -
2{\mu}_{3}(w_{3}-w_{6})X(t)\right].\label{sec_nonlinear1}
\end{eqnarray}
The choice of signs in (\ref{omega_01}) and (\ref{omega_23}) will
be motivated in the next subsection, for the moment they are just
a short-hand notation for certain non linear
combinations of the fundamental coupling constants of the theory.

It may be a bit more transparent, to put (\ref{sec_nonlinear}) and
(\ref{sec_nonlinear1}) into a matrix form and to reinsert $\Delta$:
\begin{eqnarray}\label{SecMatrix}
\frac{d}{dt}\left( \begin{matrix} R(t)\cr X(t)\end{matrix}\right)
 & = & 3\left( \begin{matrix}
2{\omega}_{1}^2 & -{\omega}_{2}^2\cr
{\omega}_{3}^2 & {\omega}_{0}^2\end{matrix}\right)\cdot \left(
\begin{matrix} u(t)\cr v(t)\end{matrix}\right)
-2\left[u(t)+\mu_3\frac{w_3-w_6}{\mu_3^2+4w_3 w_6}v(t)\right]
\left(\begin{matrix}R(t)\cr X(t)\end{matrix}\right)\cr &- &
\frac{v(t)}{\mu_3^2+4w_3 w_6}\left(\begin{matrix}0& \mu_3^2+4w_3^2\cr 
    \mu_3^2+4w_6^2&0 
\end{matrix} \right)\cdot\left(\begin{matrix}R(t)\cr X(t)\end{matrix}\right).
\end{eqnarray}

\end{widetext}
We have eight variables but only seven relations between
them. However, we still have to choose an appropriate {\em equation of
  state} ${p}=p({\rho})$. In cosmological models for late times,
$p(t)\approx 0$ is a widespread assumption.



\subsection{Rearranging the field equations into first order form}\label{Sec.VE}

For certain purposes (in particular numerical evolution) it is more
convenient to replace the 3 dynamical second order equations for the
gauge potentials by 6 first order equations for the observable
quantities $\{a,H,u,v,R,X\}$.  From
(\ref{Hubble},\ref{XX},\ref{firstcombined}) one can obtain the first
order set
\begin{eqnarray}\label{num1}
  a'(t)&=&H(t)a(t)\,,\\
  H'(t)&=&-2H^2(t)-\frac{k}{a^2(t)}+\frac{1}{3a_2}\Bigl\{{\kappa}
\left[ {\rho}(t)-3p(t)\right]\quad\nonumber\\
  &&\ \ +4\lambda_0
  -\left(a_{0}-\frac{a_{2}}{2}\right)R(t) - (b_{0}+{\sigma}_{2}) X(t)\nonumber\\
  &&\ \ +\frac{3}{4}(a_{2}-4a_{3})v^2(t)\Bigr\}\,,\label{180}\\
  u'(t)&=&u^2(t)-3H(t)u(t)+\frac{1}{3a_2}\Bigl\{{\kappa}
\left[ {\rho}(t)-3p(t)\right]\quad\nonumber\\
  &&\ \ +4 \lambda_0
  -a_{0}R(t) -(b_{0}+{\sigma}_{2}) X(t)\nonumber\\ &&\ \
-3a_{3}v^2(t)\Bigr\}\,,\\
  v'(t)&=&-\frac{1}{3} X(t)-v(t)[3H(t)-2u(t)] \,, \label{182}
\end{eqnarray}
along with Eqs.~(\ref{sec_nonlinear}, \ref{sec_nonlinear1}) for
$R'(t),X'(t)$. In addition to the dynamical geometric variables, these
equations also include the material energy density and the pressure,
$\rho(t),p(t)$.  The material energy density $\rho(t)$ is related to
the dynamical variables by (\ref{density})---a relation which could be
used to eliminate it from the system.  The energy density and pressure
are necessarily related by
\begin{equation} {\rho}^{\prime}(t) =
  -3H(t)[{\rho}(t)+p(t)].\label{rho_prime}
\end{equation} 
This relation follows, on the one hand from the basic Noether symmetry
conservation law applied to the source, and, on the other hand can be
derived directly from (\ref{density}) using the system of 6 first
order equations (\ref{sec_nonlinear}, \ref{sec_nonlinear1},
\ref{num1}--\ref{182}).

The above system needs to be supplemented by an appropriate equation
specifying a relation for $p(t)$ in terms of suitable dynamical
variables.  In GR such a relation is often taken in the form of a
fluid equation of state $p=p(\rho)$.  One could also use such an
assumption for our PG model or, more generally, one could consider any
specific relation of the form
\begin{equation}
p=p(a,H,u,v,R,X)
\end{equation}
to reduce the dynamical equations of the model to a closed system of 6
nonlinear coupled first order ordinary differential equations (1st
order ODEs) describing the dynamics of 3 geometric degrees of
freedom.  Alternatively, if one had a relation of the form
\begin{equation}
 p^{\prime}(t)  =  f[p(t), {\rho}(t), u(t), v(t), \cdots]\,, \label{p_prime}
\end{equation}
this would be sufficient for integrating the system
(\ref{sec_nonlinear}, \ref{sec_nonlinear1},
\ref{num1}--\ref{rho_prime}, \ref{p_prime}).

To investigate more general models in the presence of torsion, one
cannot just prescribe a simple equation of state. One could consider
some explicit source fields and their dynamics; these sources would
determine $\rho(t),p(t)$. Also one could relax the assumption of
vanishing source spin density.  In follow up work some systems will be
presented which might be more stable numerically.

The prescription of an (algebraic) equation of state reduces the
phase-space to one of $6$ dimensions. Systems of ODEs with algebraic
constraints are usually called {\em differential algebraic equations}
(DAEs) or also descriptor systems\footnote{See the link {\tt
    http://www4.ncsu.edu/eos/users/s/slc/www/
    RESDESCRIPT/resdescript.html}: ``Usually the term DAE refers to
  systems of ordinary differential equations $F(x', x, t)=0$ with the
  Jacobian of F with respect to $x'$ being singular.''}.  For the
numerical evaluation we would have a DAE of the form $6\oplus 1$, that
is, $6$ ODEs of first order and one algebraic equation which
makes the whole system of equations determinate.

Numerical simulations of those systems, including the case $u(t)\sim
v(t)$, will be discussed in detail in a continuation of this
paper. For the subcase of the Shie-Nester-Yo Lagrangian (\ref{SNY}),
such computations have been already done by Li, Sun, and Xi
\cite{Li:2009zzc,Li:2009gj,Ao:2010mg}.

\subsection{Acceptable choices of signs}\label{Sec.VF}

Concerning the acceptable choices of signs for the parameters: at the
Lagrangian level of analysis we know of only one necessary
requirement, namely, in order to satisfy the {\em principle of least
  action} it is absolutely necessary to take kinetic energy
terms---here meaning specifically the quadratic-in-time-derivative
terms---to have a positive coefficient.

For our model this is sufficient to fix the signs for the
quadratic-in-curvature terms, since it turns out that the scalar
curvature and pseudoscalar curvatures are each linear in the time
derivatives of certain connection coefficients.  Hence physically
in Eqs.~ (\ref{eq9:120}) one must take only the case
$\lambda_1<0,\lambda_2<0$.

Regarding the quadratic torsion terms, the situation is not so
simple. For our cosmological model at least, from (\ref{Hubble}),
(\ref{T_A}), and (\ref{RC_con}), it can be seen that only $t_0$
contains a time derivative of a gauge potential (specifically,
$R'$). Thus, by the least-action requirement on this
quadratic-in-time-derivative kinetic term one must take, from
(\ref{L_BHN'}), the coefficient $a_2<0$. Consequently, since by
convention $a_0>0$, one should require $m^+>0$.

%
Beyond these considerations, one can ascertain which constraints
are physically appropriate for the parameters only from a detailed
analysis of the equations of motion.  This is left to future work.



\section{Discussion and conclusion}\label{Sec.VI}

In this work we introduce systematically the notion of {\em even and
  odd parity terms} for the construction of gravitational field
Lagrangians in the context of the Poincar\'e gauge field theory
(PG). Exploiting the theory of {\em algebraic} invariants, a
Lagrangian results that is at most quadratic in the field strengths
torsion and curvature. Here, we rigorously include interaction terms
of even and odd parity form, like those of (vector torsion $\times$
axial vector torsion), that is, ${\cal V}\wedge\, ^{\star}{\cal A}$,
and (scalar curvature $\times$ pseudoscalar curvature), that is,
$R\wedge\, ^{\star}X$. To obtain some insight into the dynamics of
those `shadows' in a physically realistic model, we constrained
ourselves to a model containing only scalar and vector parts and their
corresponding axial versions. The $V_{\rm BHN}$ Lagrangian
(\ref{L_BHN'}) can be viewed as a generalization of the recently
presented $V_{\rm SNY'}$ Lagrangian (\ref{extendedSNY}) of Chen et
al.\ \cite{exSNY}. In light of the difficulties caused by
non-linearities \cite{Hecht:1996ay,Chen:1998ad,Yo:1999ex,Yo:2001sy},
our model (aside from adding a couple of unimportant terms quadratic
in the non-dynamic $^{(1)}T^\alpha$ torsion components) may well be
the most general PG model that can be expected to have a dynamical
connection with well behaved dynamics.

From a theoretical point of view, the inclusion of additional
interaction terms of {\em odd} parity character could explain some
empirical facts we are faced with in cosmology.  Besides the usual
handling of even parity functionals of the field variables, we treat
those of odd parity character on the same footing. This may open the
discussion to explain the empirical imbalance of matter and {\em
  antimatter} on a cosmological scale and other related questions that
are still open.

Empirically, the inclusion of additional parameters beyond those of
the model (\ref{extendedSNY}) will enhance the capacity to account for
the accelerated universe observations in terms of dynamical
geometry---dark energy could be a PG dynamical connection.  It is
noteworthy that with the new pseudoscalar cross coupling parameters
the acceleration of the universe can be more directly influenced by
the $0^-$ mode, see (\ref{180}), which is known to also couple to
fermion spin.

In this work, we present for the first time (as far as we know) the
notion of the {\em diagonalization of a Lagrangian}. This identifies
certain special parameter combinations of the primary coupling
constants that are expected to play important roles in future studies
of the dynamics of our model, and leads to the recognition of certain
conditions on the set of primary coupling constants
$\{a_{2},a_{3},w_{3},w_{6},{\mu}_{3}\}$ such that the (diagonalized)
kinetic energy matrix ${\cal T}$ has strictly positive entries.  A
working hypothesis is that these conditions are needed to have a
well-defined propagation of massive modes.

According to the diagonalization, the irreducible pieces
$^{(2)}T^{\alpha}$ and $^{(3)}T^{\alpha}$ can be associated with the
two four-vectors ${\nu}^{\mu}$ and ${\alpha}^{\mu}$ of even and odd
parity character. For these vectors to have proper evolution in time,
it is expected that certain signature properties (conjectured to be
positive) for the corresponding eigenvalues of the kinetic energy
matrix are necessary. The irreducible curvature pieces
$^{(3)}R^{\alpha\beta}$ and $^{(6)}R^{\alpha\beta}$ are essentially a
scalar $R$ and a pseudoscalar $X$, respectively.  The proper choice of
parameters is such that the associated kinetic matrix has positive
eigenvalues.

In our model, we noted that in the general PG {\em weak gravity}
sector, mediated by the coframe $\vt^\a$, the associated field
strength, the torsion, could carry modes of spin $2$, of spin $1$, and
of spin $0$ (each with even and odd parity). Restricting to the even
parity terms, this is similar to Bekenstein's TeVeS
(tensor-vector-scalar) theory \cite{BekensteinTeVeS}. However, in our
case the different modes are carried by the torsion alone. There is no
need for any other scalar or vector fields.

For {\em strong gravity,} mediated by the Lorentz connection
$\Gamma^{\a\b}$, in our model we found spin $0$ of both parities. This
restriction to zero spin modes in due to our simple Lagrangian
(\ref{L_BHN'}) in which only the scalar and pseudo-scalar pieces of
the curvature were allowed. Straightforward generalizations are
possible. In metric-affine gravity (MAG), even the inclusion of spin
$3$ modes, see \cite{Baekler:2006vw}, is possible---and all of this on
the basis of a Riemann-Cartan or metric-affine geometry of spacetime,
respectively.
\\

\begin{acknowledgments}
  We would like to thank J.~D.~Bjorken (SLAC), Milutin Blagojevi\'c
  (Belgrade), Eckehard Mielke (Mexico City), and Yuri Obukhov (London)
  for numerous helpful remarks. Advice on our DAEs by Fritz Schwarz
  (St.~Augustin) and Werner M.\ Seiler (Kassel) is gratefully
  acknowledged. One of the authors (FWH) is deeply grateful to the
  Center for Mathematics and Theoretical Physics of the National
  Central University for the hospitality extended to him in Chungli
  where this paper was initiated. JMN was supported by the National
  Science Council of the R.O.C.\ under the grants
  NSC-98-2112-M-008-00, NSC-99-2112-M-008-004, and also in part by the
  Taiwan National Center of Theoretical Sciences (NCTS). P.~Baekler
  was partly supported by the Department of Media, FH-D\"usseldorf.
\end{acknowledgments}



\end{document}